\documentclass{aa}  
\usepackage{placeins}
\usepackage{booktabs}
\usepackage{threeparttable}
\usepackage{xcolor}
\usepackage{graphicx}
\usepackage{txfonts}
\usepackage{float}
\usepackage{makecell} 

\begin{document}

\title{In-flight calibration of RADEM, the JUICE mission radiation monitor}

    \author{M. Pinto \inst{1} \and
            F. Santos \inst{1} \and
            A. Gomes \inst{1} \and
            T. M. Gonçalves \inst{1,2} \and
            L. Arruda \inst{1} \and
            P. Gonçalves \inst{1,2} \and
            L. Rodríguez-García \inst{3} \and 
            R. Vainio  \inst{4} \and
            O. Witasse \inst{5} \and 
            N. Altobelli \inst{3}
            }

   \institute{Laboratório de Instrumentação e Física Experimental de Partículas,
              Av. Prof. Gama Pinto 2, 1649-003 Lisboa, Portugal\\
              \email{mpinto@lip.pt}
              \and
              Instituto Superior Técnico - University of Lisbon, Av. Rovisco Pais 1, 1049-001 Lisboa, Portugal
              \and
              European Space Agency(ESA) European Space Astronomy Centre (ESAC) Camino bajo del Castillo, s/n Urbanización Villafranca del Castillo Villanueva de la Cañada E-28692 Madrid, Spain
              \and
              Department of Physics and Astronomy, University of Turku, FI-20014 Turku, Finland
              \and
              European Space Research and Technology Centre, European Space Agency, 2201 AZ Noordwijk, The Netherlands
             }


 
\abstract
{The RADiation-hard Electron Monitor (RADEM) aboard the JUpiter ICy moons Explorer (JUICE) mission, launched on April 14th 2023, is measuring high-energy protons and electrons during the cruise phase and will continue to do so during the nominal mission phase. However, ground calibration results were unable to explain the initial flight observations, prompting an in-flight calibration campaign.}
{Our main goal was to calibrate RADEM and develop a procedure to compute particle fluxes from the count rates obtained by the RADEM detector heads.}
{We used galactic cosmic rays (GCRs) to calibrate RADEM's sensors by increasing the respective thresholds and therefore modifying their response to high energy particles. The count rates obtained in-flight for each threshold were then compared to theoretical count rates calculated using the Baddhwar-O'Neil 2020 (BON2020) GCR model and threshold dependent response functions. These results were then used to construct a flux-reconstruction algorithm based on the bow-tie method.}
{We derived a new set of in-flight calibration coefficients for all sensors. In several cases, the in-flight calibration slopes differ by up to an order of magnitude from pre-flight ground calibration values. Proton fluxes from solar energetic particle (SEP) events, reconstructed using the bow-tie method, show good agreement (within a factor of two) with measurements from the SOlar and Heliospheric Observatory (SOHO).}
{RADEM provides accurate measurements of proton fluxes in interplanetary space and is well suited for both single-spacecraft analyses and coordinated multi-mission studies of SEPs. While electrons have been clearly identified during the JUICE Lunar-Earth gravity assis (LEGA), reconstructing their fluxes needs a more detailed analysis.}

\maketitle
%

\section{Introduction}
On April 14th, 2023, the European Space Agency (ESA) launched the  JUpiter ICy moons Explorer (JUICE) mission aboard an Arianne 5 \citep{JUICE_paper}. JUICE will be the fourth mission to orbit Jupiter, after Galileo \citep[1995-2003;][]{JupAngVar_paper}, Juno \citep[2016-ongoing;][]{Juno_paper}, and Europa Clipper \citep{clipper_paper}, launched on October 2024 and planned to arrive in 2030. Its main objective is to study Jupiter and its three icy moons, Europa, Ganymede and Callisto. The mission will characterize the subsurface oceans and potential habitability of these moons, study their surface \citep{Surface_paper} and geological activity \citep{Geophysics_paper}, and analyze the coupling between the moons and Jupiter’s magnetosphere \citep{magnetosphere_paper}. Additionally, JUICE will explore the structure and dynamics of Jupiter’s atmosphere and magnetosphere, contributing to a broader understanding of gas giant systems and the conditions that may support life \citep{Jupiter_science_paper}.

\begin{figure}[th!]
\centering
\includegraphics[width=0.95\linewidth]{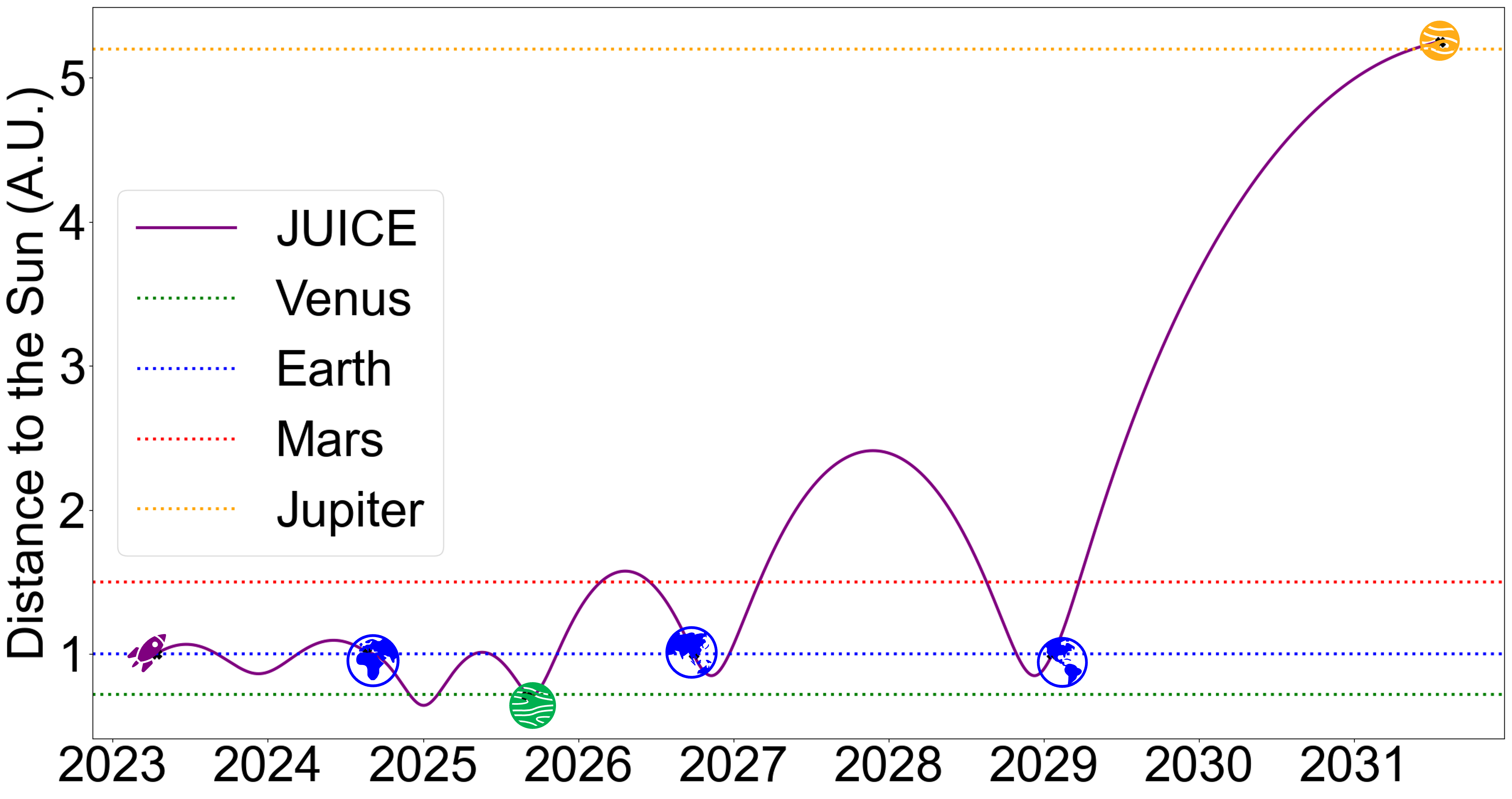}
     \caption{JUICE cruise phase trajectory (purple) as a function of its distance to the Sun and its four gravity assists. The average distance of Venus (green), Earth (blue), Mars (red) and Jupiter (orange) to the Sun are shown for reference. JUICE is expected to perform three gravity assists of Earth (the first combined with a lunar gravity assist), and one of Venus.}
     \label{fig:trajectory}
\end{figure}

JUICE will perform an eight-year cruise in interplanetary space with four gravity assist maneuvers: the first-ever Lunar-Earth gravity assist (LEGA) which took place between August 19-20, 2024, one of Venus that took place on August 31, 2025, and two of Earth planned for September 2026 and January 2029. Orbital insertion at Jupiter is scheduled for July 2031. Fig.~\ref{fig:trajectory} shows the JUICE distance to the Sun during the entire cruise phase. 

During both the cruise and nominal phases of the mission, JUICE will be exposed to a hazardous radiation environment. Energetic charged particles pose a major challenge for both crewed and robotic space missions, as they can induce long-term degradation through total ionizing dose \citep[TID;][]{Dodd2003} and displacement damage dose \citep[DDD;][]{Srour2003}. In addition, they can cause short-term disruptions such as single-event upsets, or, in extreme cases, lead to complete mission loss due to single-event latch-ups \citep{Oldham2003}. For this reason, JUICE carries the RADiation-hard Electron Monitor \citep[RADEM;][]{radem_paper,DDH_paper}. RADEM's main objective is to assist mission operations by measuring energetic electrons (0.3~MeV-10~MeV) and protons (5.0~MeV-250~MeV). In addition, it will provide unique long-term measurements of high energy electrons and protons in the Jovian system, and contribute to heliophysics science. Although RADEM was calibrated on ground, its flight measurements could not be reproduced using the results. This was attributed to inacurate calibration coefficients which led to unreliable response functions. In this work, we present the methods and results obtained during the in-flight calibration campaign of the instrument. We also describe the changes made to its configuration and the algorithm developed to compute particle fluxes. The results were validated by comparing the observations made by RADEM in September and October 2024 with the Energetic and Relativistic Nuclei and Electron \citep[ERNE;][]{ERNE_paper} aboard the SOlar and Heliospheric Observatory  \citep[SOHO;][]{SOHO_paper} when the two spacecraft were in relatively close locations.

Section \ref{sec:env} introduces the radiation environment that will be encountered by JUICE. Section \ref{sec:radem} describes RADEM in great detail. In Section \ref{sec:calib} we present the calibration campaign and the results obtained. Section \ref{sec:reconstruction} contains the methodology used to convert counts to flux and its validation. In Section \ref{sec:discussion} we summarize and discuss the main findings of the study and in Section \ref{sec:conclusions} we outline the main conclusions.

\section{The JUICE radiation environment}\label{sec:env}

The Jovian system contains large fluxes of electrons with energies up to tens of MeV. The acceleration of electrons and other charged particles within the Jovian system is driven primarily by Jupiter’s powerful and dynamic magnetosphere, which is the largest and most energetic in the Solar System \citep[e.g.][]{Kollmann2018ElectronAcceleration,Nenon2022PADs}. Particles originating from the volcanic moon Io are ionized and subsequently accelerated through various mechanisms, including radial diffusion, wave-particle interactions, and magnetic reconnection events \citep{magnetosphere_paper}. These processes are fueled by Jupiter’s rapid rotation and strong magnetic field, leading to the formation of intense radiation belts. Understanding these acceleration processes is critical for comprehending the Jovian system as a whole, to determine whether the detection of biosignatures is possible, and to ensure spacecraft protection \citep[e.g.][]{magnetosphere_paper,Nordheim2019GCREuropa, eco60}.

During its interplanetary cruise phase, JUICE will be exposed to solar energetic particles (SEPs) and galactic cosmic rays (GCRs). It will also encounter Earth's trapped radiation belts during the mission's three flybys of the planet. 
SEPs are sudden, currently unpredictable, bursts of high fluxes of ions (primarily protons but also heavier species) with energies up to GeV and electrons with energies up to $\sim$10~MeV, accelerated in solar flares or coronal mass ejections \citep[e.g.][]{Reames2013, Lario2022}. Although substantial progress has been made in SEP forecasting, including physics-based and empirical prediction models \citep[e.g.][]{Whitman2023_SEPmodels}, the precise timing, intensity, and spectral characteristics of individual events remain difficult to predict. SEP occurrence is strongly correlated with solar activity, commonly characterized by the $\sim$11-year sunspot cycle, with higher event rates during solar maximum.
JUICE was launched shortly before the solar maximum of solar cycle 25 and will remain in interplanetary space during its decreasing phase. In the last two years, RADEM has already observed tens of SEP events.

GCRs are a continuous low flux of high energy charged ions (protons and heavier ions), originating outside the Solar System \citep[e.g.][]{Potgieter2013HeliosphereGCR}. At energies up to a few GeV, their flux is modulated by transport processes in the heliosphere, including diffusion, convection with the solar wind, adiabatic cooling, and gradient and curvature drifts in the large-scale heliospheric magnetic field. As a result, GCR intensities are anti-correlated with solar activity, i.e., during periods of high solar activity the GCR flux decreases, while during solar minimum it reaches its maximum levels. Solar modulation of GCRs can be quantified by a single value known as the modulation potential, $\phi$, measured in units of MV (megavolts). $\phi$ represents the average energy loss that a cosmic ray particle experiences as it propagates through the heliosphere.
On shorter time scales, transient structures such as interplanetary coronal mass ejections can produce sudden decreases in GCR intensity, known as Forbush decreases \citep{Forbush1937}.
GCR transport in the heliosphere also results in a radial gradient. Measurements made by interplanetary missions to the outer solar system such as Voyager, Cassini and Rosetta, indicate a radial gradient of $\sim3\%/au$, that varies between solar cycles \citep{RosetaGCRrad, CassiniGCRrad}.

During the three Earth flybys, JUICE will pass through the Van Allen belts. These are regions of Earth's magnetosphere where charged particles are trapped by the geomagnetic field. They can be generally separated into inner and outer radiation belts, the former dominated by high energy protons and the latter by high energy electrons (up to $\sim$10 MeV). Earth flybys are a great opportunity to test and calibrate instruments, as the radiaditon belts are relatively well modeled and continuously monitored by missions such as the Exploration of Energization and Radiation in Geospace \citep[ERG;][]{ARASE} mission. Although JUICE has already performed the LEGA, a detailed analysis of the flyby will be published in a special issue.

\section{RADEM}\label{sec:radem}
\begin{figure}[h!]
\centering
\includegraphics[width=0.99\linewidth]{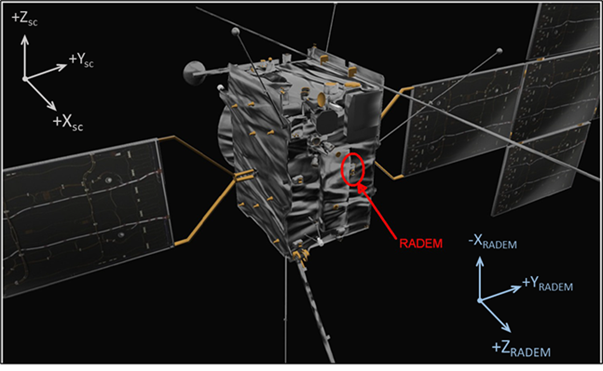}
     \caption{JUICE spacecraft visualization with marked location of RADEM (adapted from \cite{radem_paper}). Also indicated are the spacecraft and RADEM reference frames archived in the Spacecraft, Planet, Instrument, C-matrix, Events (SPICE) JUICE kernels with the names JUICE$\_$SPACECRAFT and JUICE$\_$RADEM \citep{ESA_JUICE_SPICE_2022}. The location of JUICE$\_$RADEM in the JUICE$\_$SPACECRAFT reference frame is: $X_{sc}$= 1.1853~m, $Y_{sc}$ = 0.4624~m, $Z_{sc}$ = 1.7822~m.}
     \label{fig:juice}
\end{figure}

RADEM, is a relatively low-power (3.6~W) and low mass (4.6~kg) instrument \citep{radem_paper}. It was designed to measure electrons from $\sim$~0.3 to $\sim$~40 MeV, protons from $\sim$5 to $\sim$250 MeV, and to identify ion species from helium to oxygen. It is located in the $+X_{sc}$ panel of the spacecraft in the JUICE$\_$SPACECRAFT reference frame as shown in Fig.~\ref{fig:juice}. 

\subsection{Detector Heads}\label{sec:detec}
\begin{figure}[h!]
\centering
\includegraphics[width=0.99\linewidth]{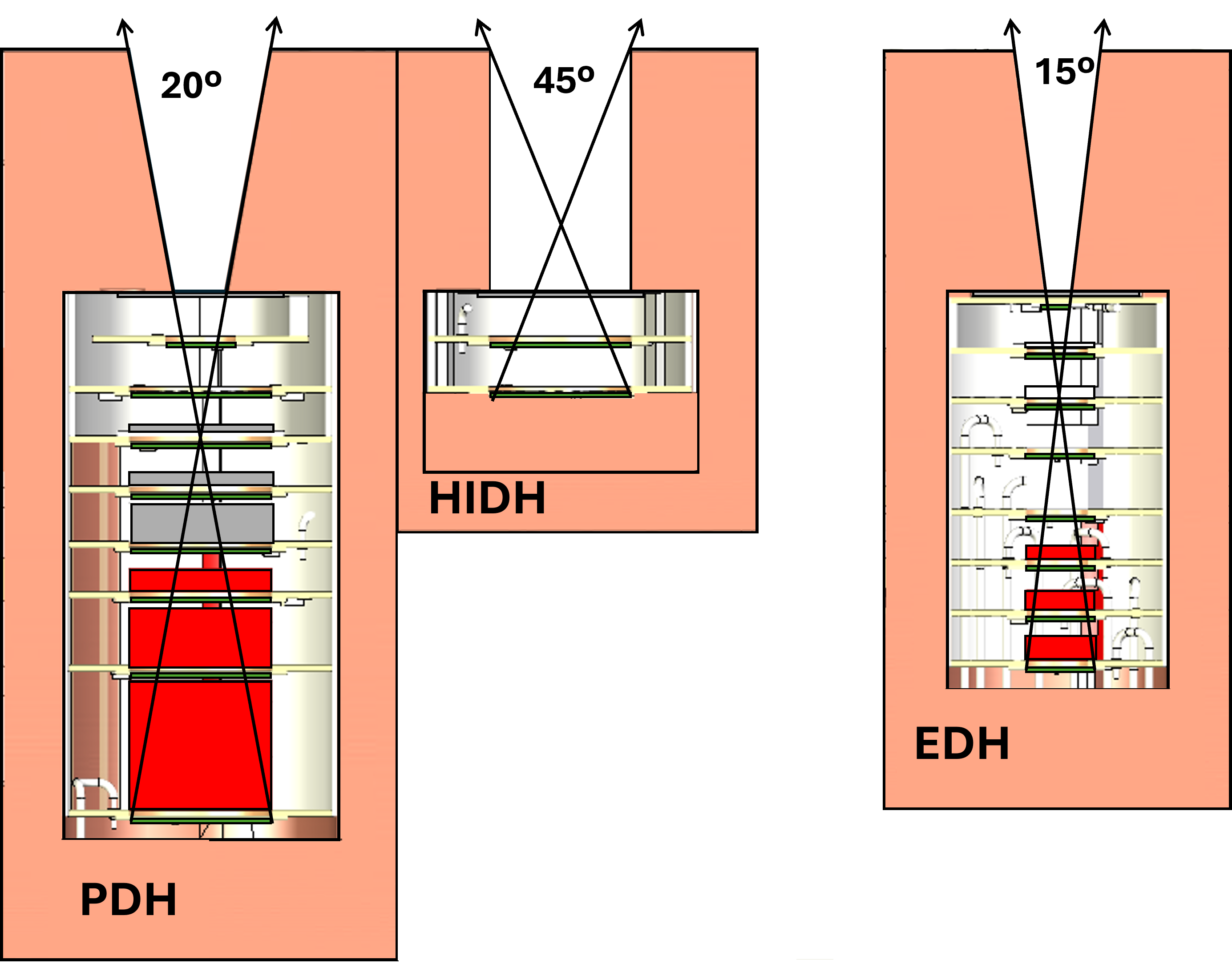}
     \caption{Side cut-view of the PDH, HIDH and EDH with their respective silicon sensors (green), copper collimators (beige), and aluminum (gray) and tantalum (red) absorbers. Dimensions are to scale.}
     \label{fig:detheads}
\end{figure}

RADEM consists of four detector heads: the Electron Detector Head (EDH), the Proton Detector Head (PDH), the Heavy Ion Detector Head (HIDH), and the Directional Detector Head (DDH). The EDH, PDH and HIDH are standard silicon-stack telescopes such as the BepiColombo Environment Radiation Monitor \citep[BERM;][]{BERM}. The EDH and PDH have eight silicon sensors each, while the HIDH only has two - see Fig.~\ref{fig:detheads}. After launch, the HIDH top sensor (HIDH~D1) became unresponsive and therefore is not considered in this analysis.

All RADEM sensors are 300 $\mu$m thick with different areas. EDH's top sensor (EDH~D1) has an area of 7.07~mm$^2$ and the other seven (EDH~D2-D8) an area of 28.3~mm$^2$. The PDH's top sensor (PDH~D1) has the same area as the EDH~D2-D8 (28.3~mm$^2$). PDH~D2-D8 and and both HIDH sensors have an area of 113~mm$^2$. The EDH, PDH and HIDH field-of-views are 15$^{\circ}$, 20$^{\circ}$, and 45$^{\circ}$, respectively. They were selected to avoid pile-up and saturation due to the large particle fluxes expected in the Jovian environment. 

The DDH is a completely novel design whose objective is to measure electron angular distributions. It incorporates 31 pixels into a single silicon wafer placed below a toroidal copper collimator. Of these pixels, 28 have a direct view of the sky, while the remaining three are fully obscured and serve as background sensors \citep{DDH_paper}. Since the DDH has a different working principle than that of the EDH, PDH and HIDH, and therefore a detailed analysis will be published separately.

More detailed information about RADEM's geometry can be found in \cite{radem_paper},\cite{DDH_paper}, and \cite{pinto2019}.

\subsection{Front-end electronics}\label{sec:fee}
Signal from all RADEM detector heads are processed in three independent IDE466 Aplication Specific Integrated Chips \citep[ASICs;][]{TimoASIC}. The three ASIC units are assigned to the DDH (DDH\_ASIC), EDH (EDH\_ASIC), and PDH and HIDH (PIDH\_ASIC), respectively. Each ASIC has 32 high-gain (HG) channels and four low-gain (LG) channels. All DDH, EDH, and PDH sensors are connected to HG channels. The two HIDH sensors are connected to LG channels.  Each channel is divided into a Voltage Analyzer (VA) and a Trigger Analyzer (TA); RADEM utilizes only the TA part since the VA's slow timing response would easily saturate in the Jovian radiation belts. The TA consists of:
\begin{enumerate}
    \item[$\bullet$] One Low Threshold (LGT) discriminator for LG channels.
    \item[$\bullet$] One Low Threshold (HGLT) and one High Threshold (HGHT) discriminator for HG channels.
\end{enumerate}

Each threshold type has a distinct dynamic range, programmable via a 10-bit Digital-to-Analog Converter (DAC). Electrical tests using external pulsers showed that the HGLT, HGHT and LGT triggers are sensitive to charges from +1.2~fC (0.027 MeV[Si]) to +0.1~pC (2.25 MeV[Si]), +15~fC (0.34 MeV[Si]) to +1 pC (22.5 MeV[Si]), and +260~fC (5.85 MeV[Si]) to +22.8~pC (513 MeV[Si]), respectively \citep{TimoASIC}.
 
Data readout is managed through 36 programmable pattern units that receive as input all 68 triggers (4 LGTs, 32 HGLTs, and 32 HGHTs). For each trigger, two options can be chosen: enable/disable (trigger is used / trigger is not used); coincidence/anti-coincidence (charge above - coincidence / below trigger - anti-coincidence). Note that not all triggers are used since there are more channels than detectors. Additional ASIC information can be found in \cite{TimoASIC}.

\begin{figure}[th!]
\centering
\includegraphics[width=0.95\linewidth]{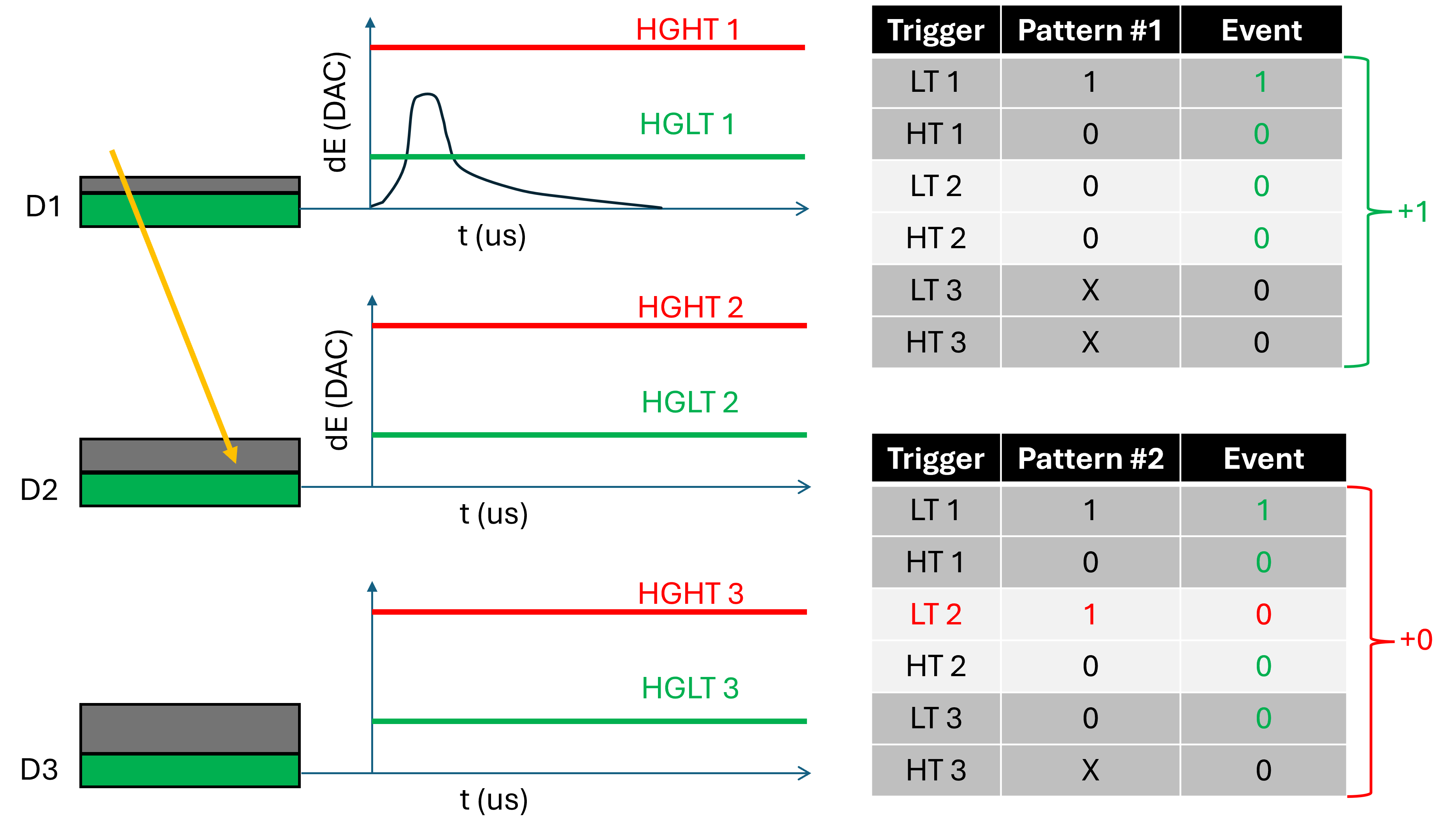}
     \caption{Example of a particle (yellow) interacting with a detector stack. The particle is able to pass through the first absorber (gray) and the first sensor (D1 - green) and stops in the second absorber. Depending on the the particle species (electron, proton or ion), and on the threshold of HGLT~1 and HGHT~1, it generates a signal (black) that may or not activate the first pattern/counter but not the second.}
     \label{fig:ASIC}
\end{figure}

Figure~\ref{fig:ASIC} shows an example of an event produced by the passage of a particle processed by two different coincidence patterns. In both coincidence patterns, the HGLT~1 is in coincidence and its value is such that electrons do not deposit enough energy to trigger it. Additionally, the HGHT~1 is in anti-coincidence so that helium and heavier ions are not detected. In the first coincidence pattern the HGLT~2 is in anti-coincidence while all other triggers deactivated. This ensures that particles reaching the second or subsequent sensors are not counted. In the second coincidence pattern, the HGLT~2 and HGHT~2 have the same state as the HGLT~1 and HGHT~2 respectively. The HGLT~3 is in anti-coincidence while the HGHT~3 is not used. The first pattern was chosen to count protons that deposit energy in the first detector but not in the second. The second pattern counts protons only if they deposit energy in the first and second detectors and therefore have higher energy, but not in the third. In yellow, a proton goes through the first sensor (D1 - green) and stops in the absorber (gray). As a consequence, it generates a signal (black) in the first sensor (D1) connected to HG~1, but not in sensors D2 and D3, each connected to HG~2 and HG~3, respectively. If the signal amplitude is between HGLT~1 and HGHT~1 as shown, then the counter associated with the first pattern increases by one while the second pattern remains the same. If the signal is below HGLT~1 or above HGHT~1, then none of the counters will increment. This is the detection principle of RADEM. Depending on the threshold values and the pattern settings, the detected particle species and energies can be completely changed. In fact, by adjusting the thresholds, the EDH can act as a proton detector, and the PDH can measure electrons.

To properly characterize the detector and separate particle species, it is necessary to find the correspondence between the trigger thresholds in DAC and energy (MeV). This is the main motivation behind the present work, as the calibration coefficients obtained from ground calibration do not allow to reproduce the flight data \citep{radem_paper}.

\subsection{Detection bins}\label{sec:bins}
Each RADEM coincidence pattern is connected to one of 108 counters (36 per ASIC). However, only a subset of these registers is sent to Earth. More specifically, 31 counters from the DDH\_ASIC, 18 from the EDH\_ASIC, and 17 from the PIDH\_ASIC are read periodically (default value is 60s) and sent as telemetry. These counters are processed individually, except for five counters from the EDH\_ASIC and PIDH\_ASIC, which are summed before being sent. We refer to each counter or group of counters as detection bins. 

The detection bins are grouped into five categories in the RADEM raw data accessible through the ESA Planetary Science Archive (www.psa.int). These are: PROTONS (eight detection bins from the PIDH\_ASIC); ELECTRONS (eight detection bins from the EDH\_ASIC); HEAVY\_IONS  (two detection bins from the PIDH\_ASIC); DD  (31 detection bins from the DD\_ASIC); and CUSTOM  (12 detection bins: 1-5 and 12 from the EDH\_ASIC and 6-11 from the PIDH\_ASIC). The rationale behind the CUSTOM detection bins is that they can be frequently reconfigured to serve different purposes while keeping the others relatively stable. 

The count rate in each detection bin is a function of the environment particle flux, $J(E)$, and the response functions, $RF_i(E)$, where $i$ is the particle type (electrons, protons, helium, etc):

\begin{equation}\label{eq:CountRates}
    R = \sum_i\int J_i(E)RF_i(E)dE 
\end{equation}

As described in Section \ref{sec:fee}, by changing the threshold values and pattern settings, defined in this manuscript as configurations, the response function RF(E) of each detection bin can be adapted to measure specific particle species and energies. Up to the date of writing, RADEM had two main science configurations. These are discussed in detail in Sect.~\ref{sec:reconstruction} and their ASIC settings are given in Appendix~\ref{app:config1}~and~\ref{app:config2}. Several other configurations were used for specific purposes, such as the Near Earth Commission Phase but are outside the scope of this paper.

\section{In-flight calibration}\label{sec:calib}
Calibration of silicon stack detectors, such as the EDH, PDH and HIDH, is usually performed with multiple electron, proton, and ion beams at different energies and/or with radioactive sources. The goal is to measure the deposited energy in each detector and compare the experimentally obtained values (in electrical units) with Monte Carlo simulations of the interactions between the beams and the instrument \citep[e.g.][]{BERM}. Although ground calibration was done for the EDH and PDH (DDH and HIDH calibration was limited) and the results published in \cite{radem_paper}, they could not be used to reproduce the flight measurements. 
This can be due to a variety of reasons. For example, \cite{radem_paper} assumed that zero DAC value corresponds to a zero energy threshold—that is, the calibration curve is taken to have a zero intercept (offset). However, the ASIC specification \citep{TimoASIC} explicitly defines a dedicated, non-zero offset term. This offset represents the threshold energy when the DAC value is zero and has a large effect on the DAC-to-MeV conversion. Other reasons include electrical integration in the spacecraft increasing or decreasing noise; and temperature effects on the electrical properties of the detector and ASIC (thermal noise, gain, etc).

For this reason, we developed a protocol to calibrate each RADEM trigger threshold in interplanetary space using GCRs. By independently scanning the threshold of the triggers connected to each sensor, we varied the deposited energy required for particle detection and, consequently, the corresponding count rate. The resulting flight data was then compared with detailed Monte Carlo simulations to obtain the calibration coefficients for each trigger.

\subsection{Simulation}

\subsubsection{Calibration Response Functions}\label{sec:calibRFs}
\begin{figure}[th!]
\centering
\includegraphics[width=0.95\linewidth]{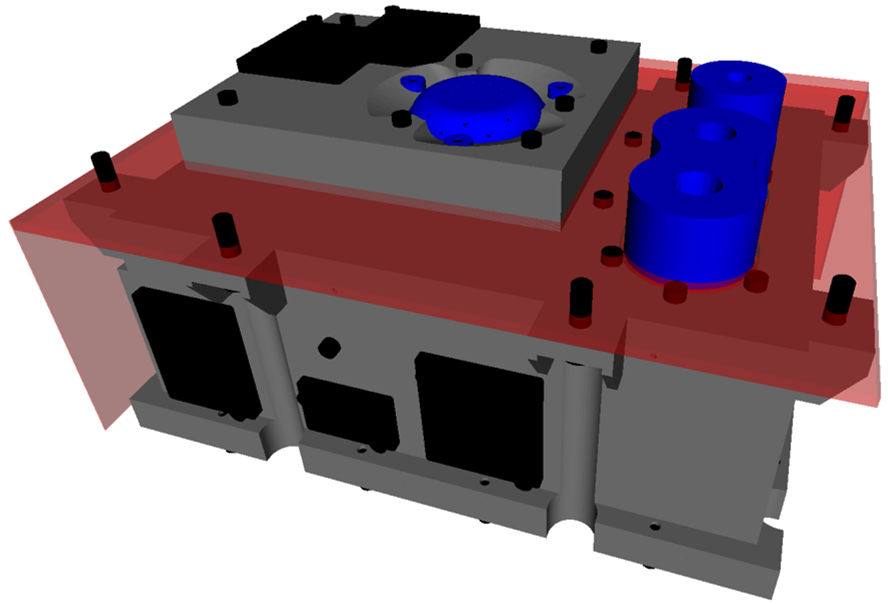}
     \caption{RADEM Geant4 model. The aluminum is depicted in gray, the tantalum shielding in black, the copper collimator in blue, and the 6-wall spacecraft aluminum equivalent in red.}
     \label{fig:GEANT4}
\end{figure}

To characterize RADEM's response to particles, we computed its response functions using the Geometry and Tracking 4 \citep[Geant4;][]{Geant4} simulations with the method described in \cite{DDH_paper}. We simulated protons, helium, carbon, nitrogen, and oxygen particles (the most abundant GCR species) isotropically with energies from 1 MeV/nucleon to 20 GeV/nucleon - the GCR flux peaks at energies close to a few hundreds of MeV/nucleon \citep{Potgieter2013HeliosphereGCR,BON2020}. Figure~\ref{fig:GEANT4} shows the geometry of the simulation. It should be noted that the spacecraft shielding was approximated as a 6-sided aluminum wall following the recommendations provided by the JUICE manufacturer \citep{pinto2019}. While this is not a detailed model and one of the main sources of uncertainty (over and/or underestimating shielding for particles entering from certain angles), it gives a first order approximation with minimal computational cost.

\begin{figure}[th!]
\centering
\includegraphics[width=0.99\linewidth]{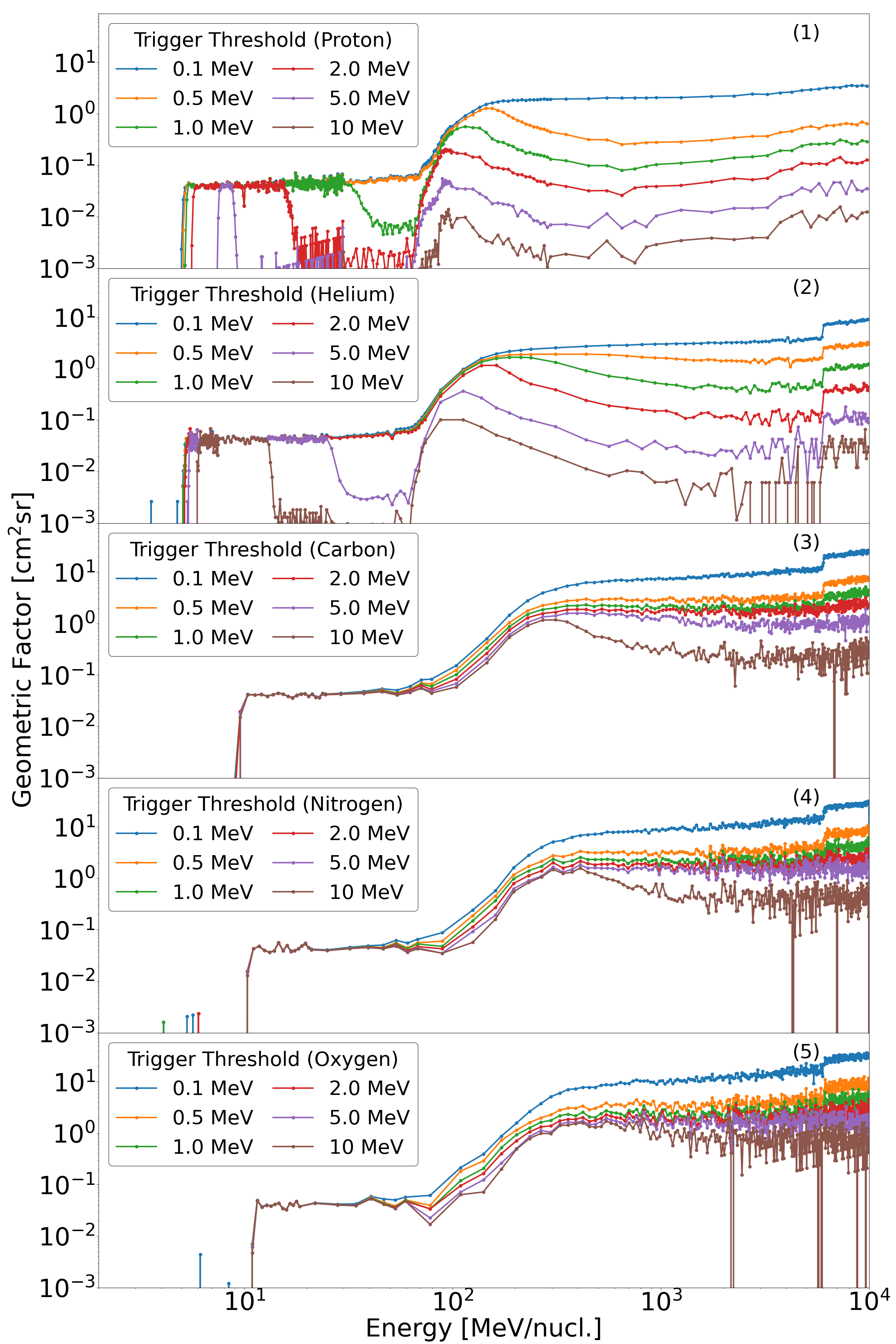}
     \caption{Proton (1), helium (2), carbon (3), nitrogen (4), and oxygen (5) response functions of the first PDH sensor (D1). Each color corresponds to a different deposited energy threshold. The larger the threshold, the less efficient the detector is for high energy proton, helium, and carbon nuclei. Efficiency to lower energy particles decreases for proton and helium nuclei if the threshold is larger than 0.5~MeV and 2-5~MeV, respectively. Efficiency of low energy heavy ions (Z>2) only decreases at very high thresholds}
     \label{fig:ThRFsP} 
\end{figure}

Figure~\ref{fig:ThRFsP} shows the response functions of the first PDH sensor (D1) to proton, helium, carbon, nitrogen, and oxygen nuclei for different thresholds from 0.1 to 10 MeV \citep[meaningful range for the HGLT and HGHT triggers according to][]{TimoASIC}. As the threshold increases, the sensitivity range to protons with energies below 70~MeV decreases in width but not in magnitude. This happens because these protons enter through the collimator aperture almost perpendicularly to the sensor surface. For low energy thresholds, all incident protons are detected, whereas at higher thresholds, only those stopping in the sensor are registered. If the threshold exceeds the incident proton energy, no detection occurs. This effect is less noticeable for helium nuclei in the same energy range since their stopping power is higher. For heavy ions (carbon, nitrogen, and oxygen), all nuclei with energies below 70~MeV/nucleon are detected assuming these thresholds since their stopping power in silicon is much higher than that of protons and exceeds the selected thresholds.

Particles with energies above 70 MeV/nucl. can penetrate the collimator. This can be seen in the response functions as an increase in the geometric factor. At these energies, increasing the threshold decreases the geometric factor. They coincide with the peak flux of GCRs (few hundreds of MeV/nucl.) and therefore are the target of our calibration. 

\subsubsection{GCR flux}\label{sec:GCRflux}
\begin{figure}[th!]
\centering
\includegraphics[width=0.99\linewidth]{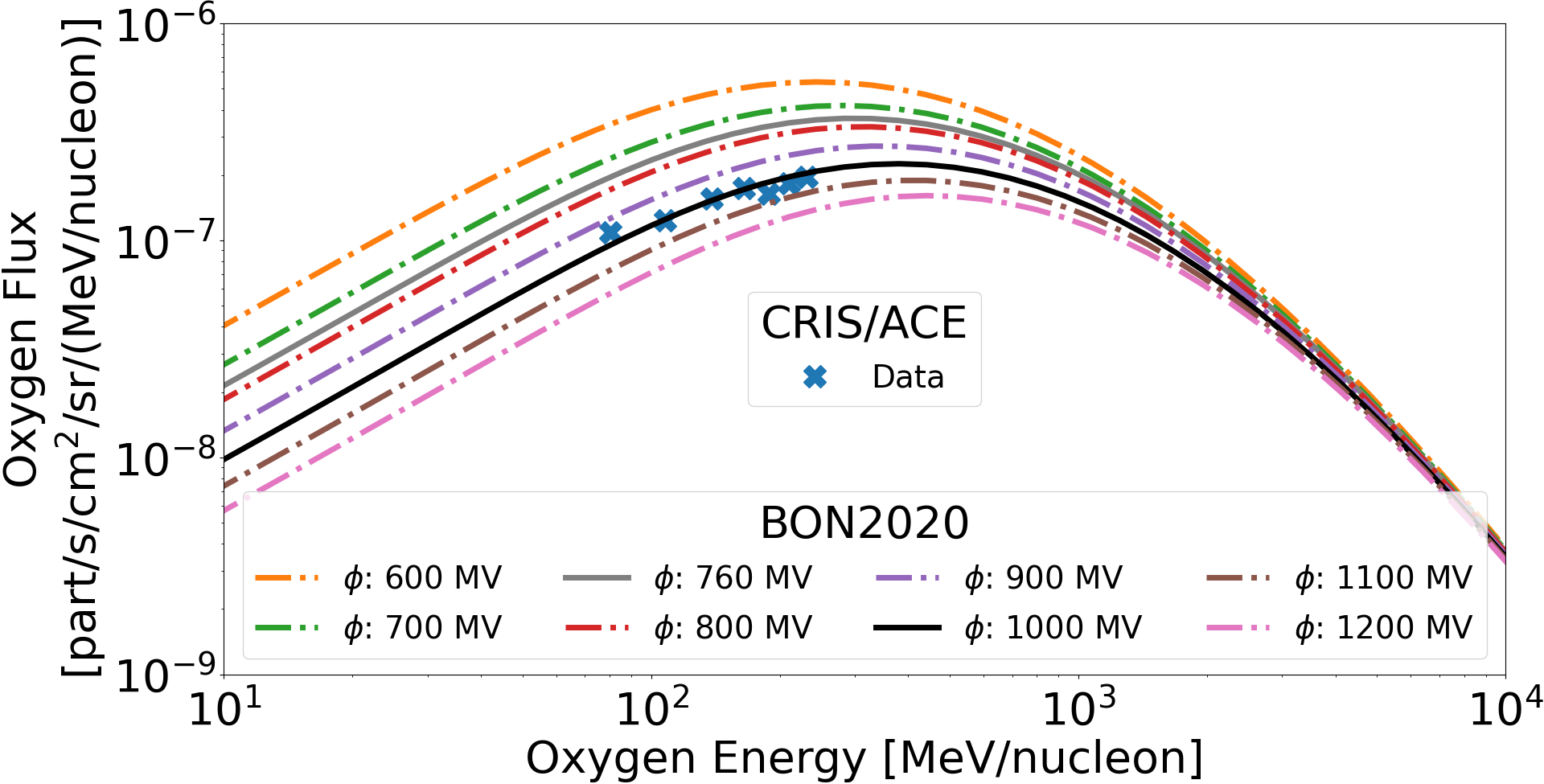}
     \caption{GCR oxygen differential flux computed by the BON2020 model (dotted dashed lines) for different modulation potentials. The gray and black lines show the spectra obtained using the modulation potentials calculated using neutron monitor and CRIS data for May 2024 (RADEM calibration campaign period), respectively. The ACE/CRIS oxygen flux measurements during the same period are shown as blue points. A modulation factor of 1000 MV was found to be in much better agreement with flight observations.}
     \label{fig:BONvsACE}
\end{figure}

Since no space detector currently covers all GCR particle species and energies, we used the Badhwar-O'Neill \citep[BON2020;][]{BON2020} GCR model to obtain the differential fluxes of protons, helium, carbon, nitrogen, and oxygen at 1 astronomical unit (au). BON2020's only free parameter is the modulation potential, $\phi$. To determine $\phi$ at a given time, \cite{BON2020} uses the integral oxygen flux measured by the Cosmic Ray Isotope Spectrometer \citep[CRIS;][]{Stone1998CRIS} aboard the Advanced Composition Explorer \citep[ACE;][]{ace} orbiting the L1 libration point. By applying this procedure to data acquired by CRIS during the RADEM calibration campaign (20--27 May 2024), described in Sect.~\ref{sec:calib}, we obtained a modulation potential of $\sim$1000 MV. Other techniques using ground neutron monitor data, such as \cite{usoskin2017heliospheric}, give a $\phi$ of 760 MV for the same time period. However, as shown in Fig.~\ref{fig:BONvsACE}, a $\phi$ of 1000 MV provides better agreement between the BON2020 flux and flight data, and therefore was used for the rest of this paper.

During the calibration interval, neutron monitor data from the Oulu station indicate GCR intensity variations of approximately 2\%, with no pronounced Forbush decreases. Concurrent in-situ solar wind measurements from STEREO-A show solar wind speeds between 300--450 km s$^{-1}$ and magnetic field magnitudes between 5--15 nT, consistent with nominal slow solar wind conditions and the absence of strong transient magnetic obstacles. These observations support the assumption of a time-static modulation potential during the calibration period. Assuming linear propagation of flux variations into the calibration coefficients, a 2\% variation in GCR intensity would translate into a comparable change in the derived calibration factor, which remains below the statistical uncertainty of the fit. The impact of short-term heliospheric variability on the calibration is therefore considered negligible.

Given that JUICE was at 1 au during the calibration campaign, we did not apply any radial correction.

\subsubsection{Simulated count rates}
\begin{figure}[th!]
\centering
\includegraphics[width=0.99\linewidth]{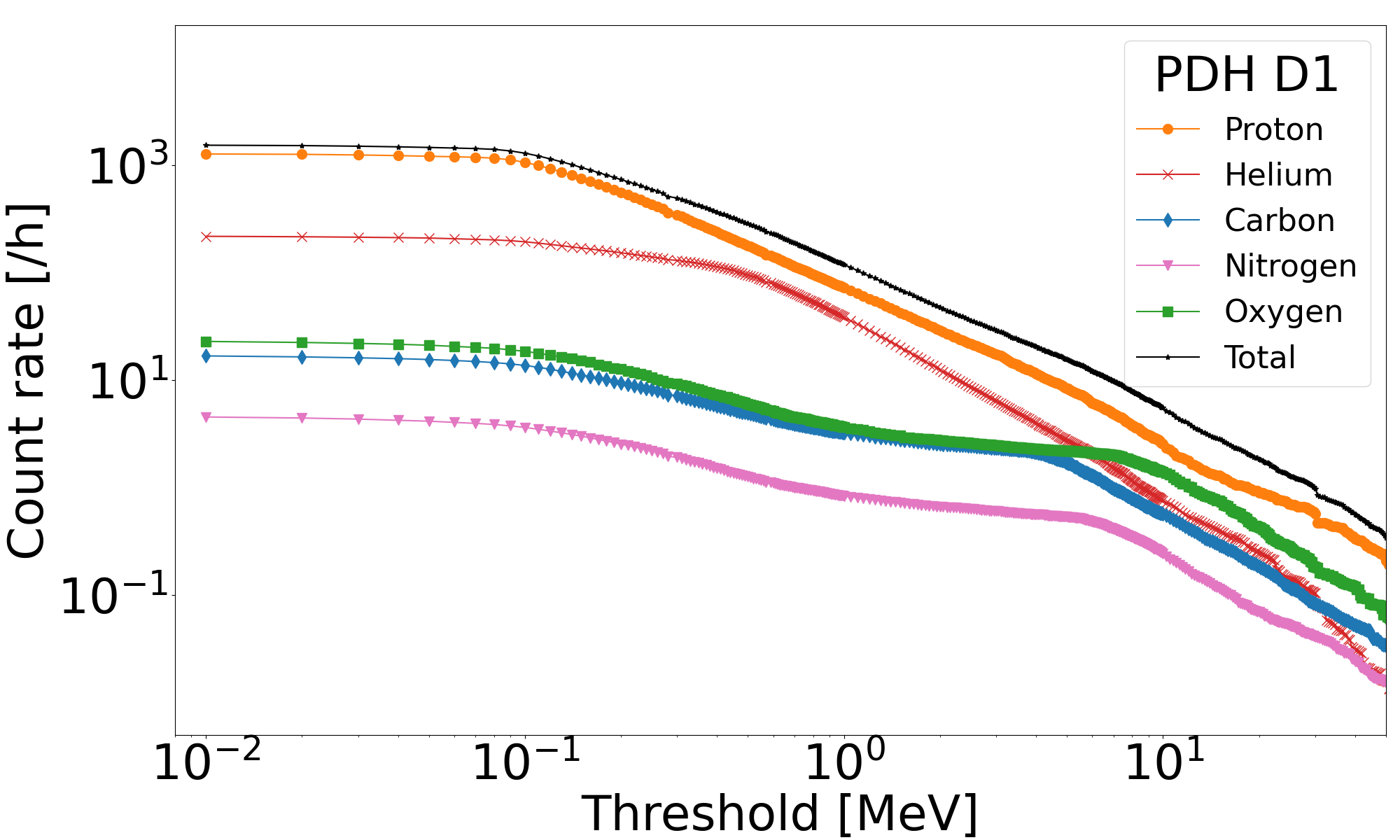}
     \caption{Proton (green), helium (red), carbon (blue), nitrogen (pink), oxygen(green), and total simulated count rates for the first PDH sensor as a function of threshold. Protons account for more than 50~\% of the counts independently of the threshold, but the combined counts from the other ions are non-negligible.}
     \label{fig:SimCRates}
\end{figure}

The count rates as a function of the energy thresholds were calculated using Eq.~\ref{eq:CountRates}, based on the response functions obtained in Sect.~\ref{sec:calibRFs} and the GCR flux computed in Sect.~\ref{sec:GCRflux}. Figure~\ref{fig:SimCRates} shows the count rates for protons, helium, carbon, nitrogen, and oxygen for the first PDH sensor as a function of the threshold. While the proton count rate dominates across all thresholds, other ions contribute up to $\sim$40\% of the counts at the considered thresholds. Heavier ions such as Fe are much less abundant, and the effect of the threshold is similar to that of the CNO group in this threshold range  \citep{Potgieter2013HeliosphereGCR,BON2020}. For this reason, ions heavier than oxygen were not considered in the calibration.

\subsection{Calibration campaign}
The calibration campaign was performed between the 21st (08:00 UT) and the 27th (23:16 UT) of May 2024. It was executed following the Mother's Day events \citep{motherdaySEPevent2024} to avoid contamination from associated SEPs. Since the number of downlinked detection bins is limited, and the number of triggers is 2x the number of sensors (with the exception of the HIDH - Sect.~\ref{sec:fee}), the campaign was split into two intervals. The integration times increased with the threshold since higher thresholds lead to significantly lower count rates. The integration times also varied between sensors due to differences in their areas and positions within the instrument. 

\begin{figure*}[]
\centering
\includegraphics[width=0.99\linewidth]{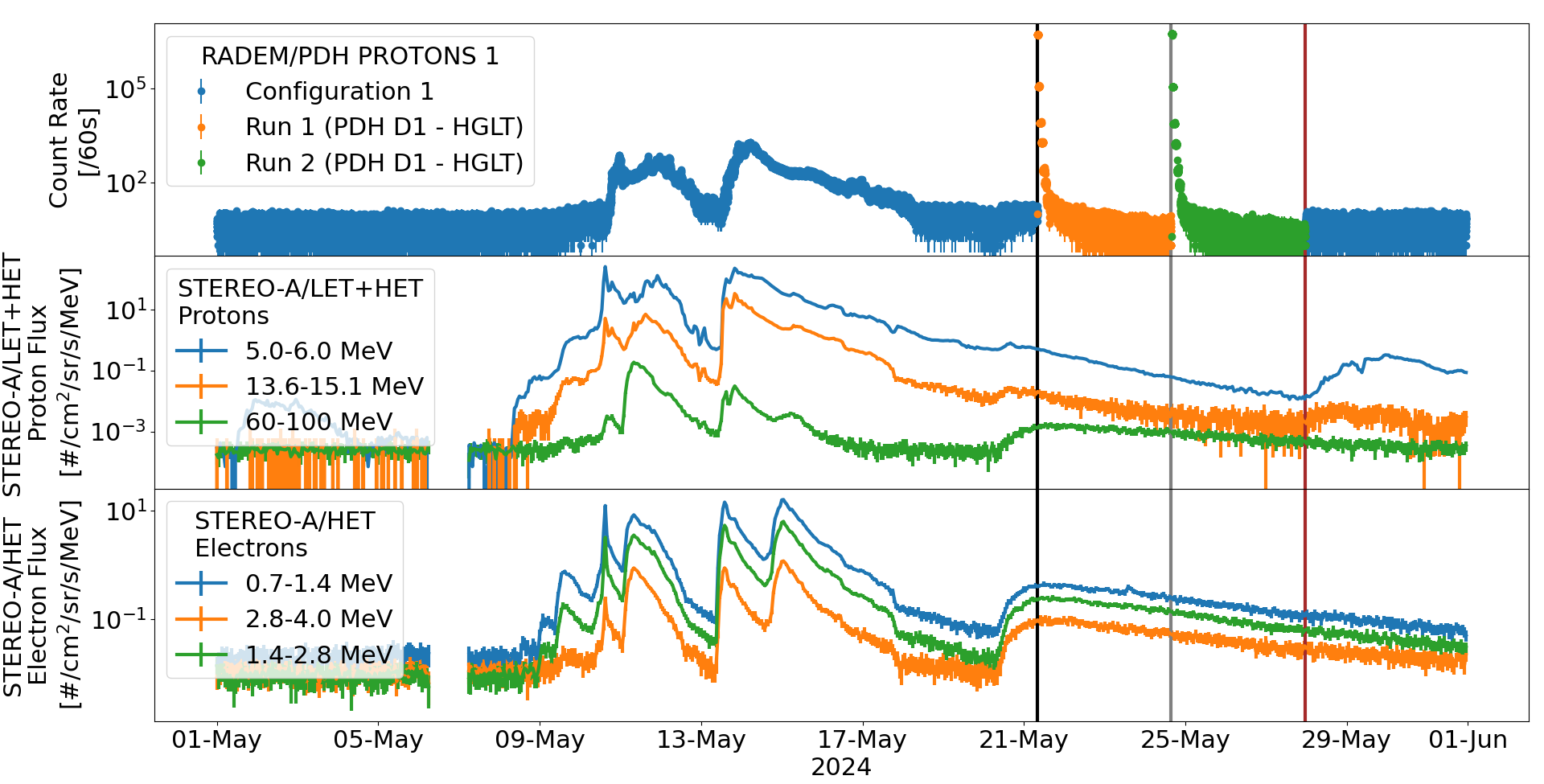}
     \caption{Count rates in the RADEM PROTONS 1 detection bin (top), and proton (middle) and electron (bottom) fluxes for selected energies measured by the STEREO-A LET (protons only) and HET instruments in May 2024. Vertical lines represent the beginning of Run 1 (black), the start of Run 2 (gray), and the end of the calibration scan (red). The colors in each panel correspond simply to different detections and have no relation between panels.}
     \label{fig:MayEvtHET}
\end{figure*}

The top panel of Fig.~\ref{fig:MayEvtHET} shows the data acquired in May 2024 in the PROTONS~1 detection bin. During the two runs (orange and green), this bin counted particles depositing energy in the first PDH sensor above its HGLT trigger. The count rate decreased by several orders of magnitude during both runs. At the peak, corresponding to very low thresholds, the count rate was much higher than the simulation estimations without thresholds. This can be attributed to electronic noise. Additionally, it can be seen that at the beginning of Run~1, the rate in PROTONS~1 had not yet returned to the background level observed before the 9th of May and after the 27th of May. This was associated with a SEP event that hit JUICE on the 20th of May and that resulted in excess counts of up to a factor of 2 during Run~1 in relation to Run~2. Particles from this event contaminated the calibration campaign by increasing the total count rate observed during the calibration campaign. Since it is not possible to accurately disentagle the counts from both particle populations (SEPs and GCRs), the contribution from SEPs had to be assessed.

\subsubsection{Impact of the 20 May SEP event}
\begin{figure}[]
\centering
\includegraphics[width=0.99\linewidth]{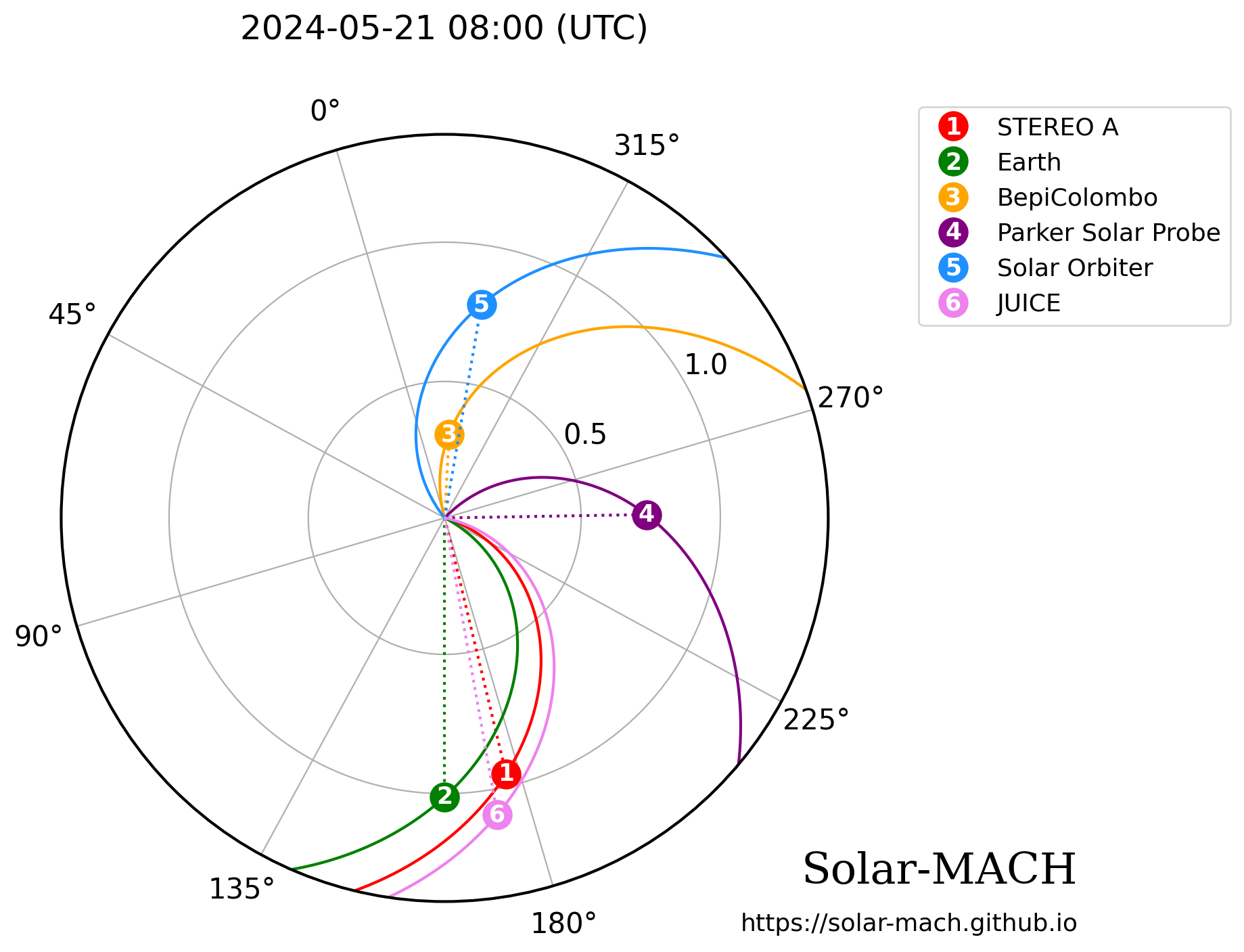}
     \caption{Spatial distribution of interplanetary spacecraft and their nominal magnetic connectivity at the start of the calibration scan - 08:00 UT on May 21, 2024. The spacecraft constellation was produced using the Solar-MACH tool \citep{SolarMach}, which is accessible online at https://doi.org/10.5281/zenodo.7016783. As shown, JUICE (pink) was relatively close to STEREO-A (red).}
     \label{fig:SolarMach}
\end{figure}

During the 20 May SEP event, JUICE was separated from STEREO-A by less than 0.13 au and $\sim6^{\circ}$ in magnetic footpoint longitude (calculated assuming solar wind speed of 400~km~s$^{-1}$) as shown in Fig.~\ref{fig:SolarMach}. The Solar TErrestrial RElations Observatory \citep[STEREO-A;][]{STEREOAref}'s High Energy Telescope \citep[HET;][]{vonRosenvinge2008} observed both protons and electrons during the entire event as shown in Fig.~\ref{fig:MayEvtHET}. The peak flux was close to the start of the calibration scan, indicating that the contribution of this SEP event to the total counts decreased over time. STEREO-A's Low Energy Telescope \citep[LET;][]{Mewaldt2008_STEREO_LET} also measured the proton flux in RADEM's lower energy range (5-12~MeV).

To compute the contribution of SEPs to the RADEM count rate, we fitted LET and HET data for protons and HET data for electrons at the beginning of Run~1 and Run~2 to a rolling power-law: 

\begin{equation}
    J(E) = J_0 E^{-\gamma} e^{-E/E_r} 
    \label{eq:rol_power_function}
\end{equation}

where $J$ is the particle flux at a given energy, E, $J_0$ is a normalization factor, $\gamma$ is the spectral index, and $E_r$ is the rolling energy. $J(E)$ was then inserted to Eq.~\ref{eq:CountRates} to calculate the count rates. At the beginning of Run~1, SEP protons dominate the count rate over GCRs for thresholds from about 1.6 to 5.2~MeV, which should correspond to high DAC values for the HGLT triggers and low DAC values for the HGHT triggers \citep{TimoASIC}. When Run~2 started, counts for GCRs accounted for at least 90\% of the counts. Electron contribution to the total count rate was negligible in both cases. This explains the differences in the count rates between the two runs. It also means that data from Run~2 can be used without restrictions, while data from Run~1 is contaminated. 

Although it may be tempting to use STEREO-A measurements to reconstruct the RADEM count rate for each acquisition period and determine its precise contribution, the particle populations sampled by the two instruments are not necessarily the same. HET points towards the Sun via the Parker spiral, whereas RADEM points away from the Sun. LET measures the proton flux in 16 directions but in a lower energy range (\textless 12~MeV). Furthermore, the two spacecrafts are not exactly at the same location. For these reasons, STEREO-A data are only used to evaluate contamination during calibration, but not to calibrate it.

\subsubsection{Calibration curves}
Since contamination from the May 20th event was very small during Run 2, calibration was performed whenever possible with data from this run only. However, eight HGLT triggers were not scanned during the second Run and therefore had to be calibrated using Run~1 data (EDH HGLT4-6 and PDH HGLT4-8). 

All HGLT triggers showed significant electronic noise at low thresholds, with count rates several orders of magnitude higher than the maximum estimated by simulations. To remove the noise, data points with count rates exceeding 10 times the simulated maximum were fitted to a line. The resulting fit was then used to calculate the signal-to-noise ratio for each threshold. Thresholds with a signal-to-noise ratio below 10 were discarded.

\begin{figure}[h!]
\centering
\includegraphics[width=0.99\linewidth]{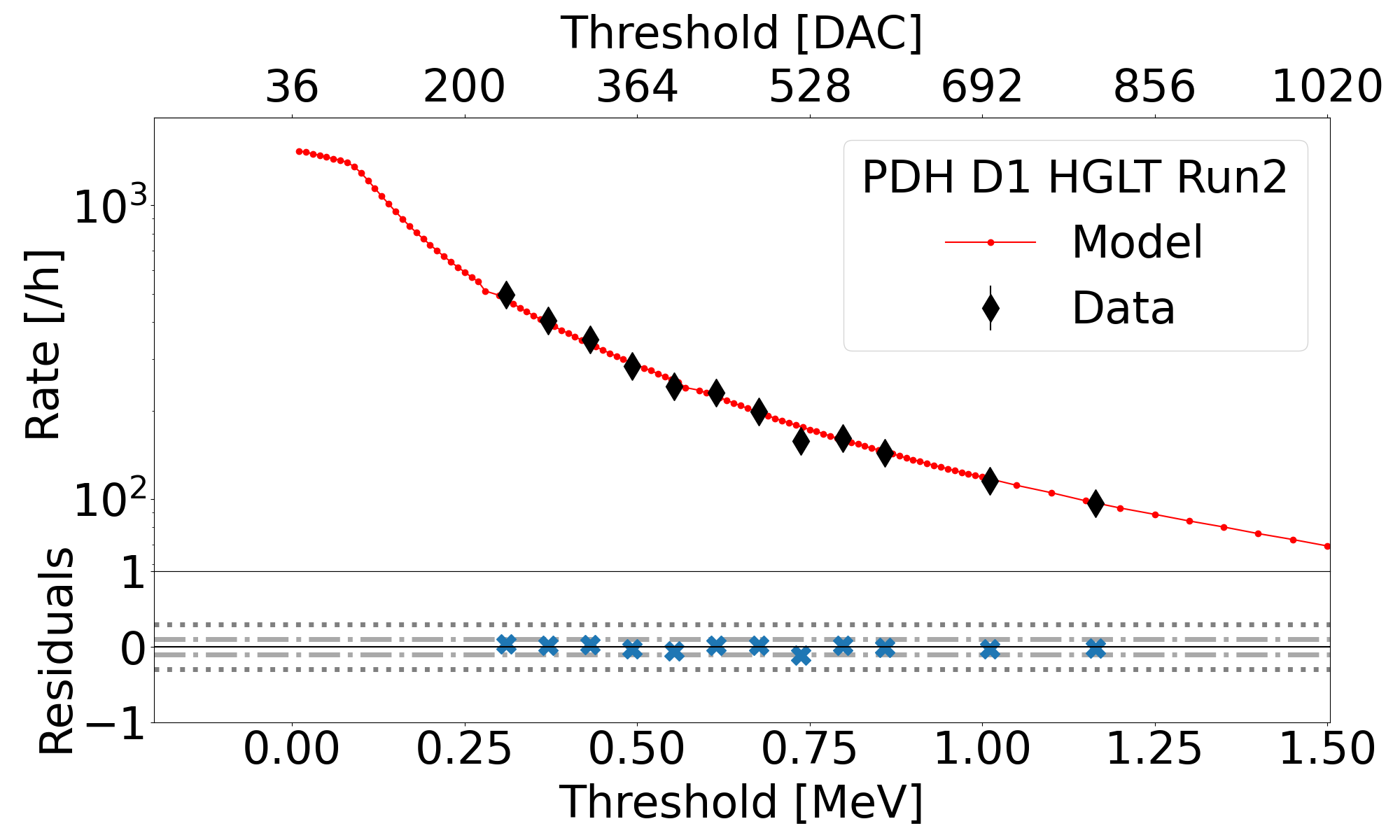}
     \caption{The top panel shows the comparison between the flight Run 2 (black) and simulated count rates (red) for the PDH~D1~HGHL trigger after applying the calibration coefficients. The bottom panel shows the residuals in blue, with the dotted, dashed, and dotted lines representing a deviation of 10 and 30\% respectively.}
     \label{fig:fit}
\end{figure}

After removing the noise, we determined the threshold in MeV for each DAC by finding the simulated count rates that matched the flight measurements. The resulting pairs of MeV and DAC thresholds were then fitted to a line to retrieve the slope and intercept. Fig.~\ref{fig:fit} shows the comparison between flight and simulated data for the PDH~D1~HGLT after applying the conversion factors. The two are in very good agreement, with most data points within 10\%.

The results for all triggers (with the exception of the DDH that will be published independently) are shown in Table~\ref{table:CalFactors}. The slope was found to be within $\sim$20\% across all HGLT and HGHT triggers of both PDH and EDH. HGHT triggers exhibit slopes $\sim$7-10 times larger than those of the HGLT, as expected given their different dynamic ranges \citep{TimoASIC}. The calibration coefficients of the 7th and 8th PDH and the 8th EDH sensor triggers had higher slopes than the others. This is likely due to poor modeling of the spacecraft shielding, which primarily affects the bottom of the detector heads - more detail is given in Sect.~\ref{sec:calibRFs}. The intersect has much higher variability ($\sim$40\%), likely due to a combination of electronic noise and a limited number of data points at very low thresholds. Negative values indicate that the pedestal is above 0 DAC, whereas positive values correspond to an offset consistent with the ASIC electrical characterization \citep{TimoASIC}.

\begin{table*}[]
\centering
\caption{Calibration coefficients of each trigger in the EDH, PDH, and HIDH. Run~2 data was used when available.}
\begin{tabular}{cccccc}
\hline
Detector Head             & Sensor \#  & 
Trigger & Run & m [MeV/DAC] & b [DAC]\\ \hline

EDH	&	1	&	HGLT	&	2	&	1.72E-03 $\pm$ 9.26E-05	&	-0.08 $\pm$ 0.05	\\
EDH	&	2	&	HGLT	&	2	&	1.74E-03 $\pm$ 2.99E-05	&	-0.10 $\pm$ 0.01	\\
EDH	&	3	&	HGLT	&	2	&	1.72E-03 $\pm$ 2.58E-05	&	-0.08 $\pm$ 0.01	\\
EDH	&	4	&	HGLT	&	1	&	1.70E-03 $\pm$ 2.86E-05	&	-0.20 $\pm$ 0.01	\\
EDH	&	5	&	HGLT	&	1	&	1.62E-03 $\pm$ 2.05E-05	&	-0.15 $\pm$ 0.01	\\
EDH	&	6	&	HGLT	&	1	&	1.64E-03 $\pm$ 2.96E-05	&	-0.18 $\pm$ 0.01	\\
EDH	&	7	&	HGLT	&	2	&	1.72E-03 $\pm$ 2.14E-05	&	-0.10 $\pm$ 0.01	\\
EDH	&	8	&	HGLT	&	2	&	1.88E-03 $\pm$ 2.79E-05	&	-0.13 $\pm$ 0.01	\\\hline
EDH	&	1	&	HGHT	&	2	&	1.92E-02 $\pm$ 2.04E-03	&	-0.21 $\pm$ 0.61	\\
EDH	&	2	&	HGHT	&	2	&	1.60E-02 $\pm$ 7.67E-04	&	0.22 $\pm$ 0.23	\\
EDH	&	3	&	HGHT	&	2	&	1.80E-02 $\pm$ 7.39E-04	&	-0.08 $\pm$ 0.22	\\
EDH	&	4	&	HGHT	&	2	&	1.64E-02 $\pm$ 6.62E-04	&	0.07 $\pm$ 0.20	\\
EDH	&	5	&	HGHT	&	2	&	1.74E-02 $\pm$ 6.22E-04	&	0.07 $\pm$ 0.18	\\
EDH	&	6	&	HGHT	&	2	&	1.77E-02 $\pm$ 6.29E-04	&	-0.22 $\pm$ 0.19	\\
EDH	&	7	&	HGHT	&	2	&	1.68E-02 $\pm$ 4.16E-04	&	0.05 $\pm$ 0.12	\\
EDH	&	8	&	HGHT	&	2	&	1.87E-02 $\pm$ 7.39E-04	&	-0.03 $\pm$ 0.22	\\\hline\hline
PDH	&	1	&	HGLT	&	2	&	1.52E-03 $\pm$ 4.08E-05	&	-0.06 $\pm$ 0.02	\\
PDH	&	2	&	HGLT	&	2	&	1.83E-03 $\pm$ 3.36E-05	&	-0.16 $\pm$ 0.02	\\
PDH	&	3	&	HGLT	&	2	&	1.83E-03 $\pm$ 2.41E-05	&	-0.16 $\pm$ 0.01	\\
PDH	&	4	&	HGLT	&	1	&	1.80E-03 $\pm$ 3.68E-05	&	-0.27 $\pm$ 0.02	\\
PDH	&	5	&	HGLT	&	1	&	1.77E-03 $\pm$ 3.60E-05	&	-0.26 $\pm$ 0.02	\\
PDH	&	6	&	HGLT	&	1	&	1.78E-03 $\pm$ 4.33E-05	&	-0.27 $\pm$ 0.02	\\
PDH	&	7	&	HGLT	&	1	&	2.01E-03 $\pm$ 4.33E-05	&	-0.19 $\pm$ 0.02	\\
PDH	&	8	&	HGLT	&	1	&	2.52E-03 $\pm$ 6.24E-05	&	-0.30 $\pm$ 0.04	\\\hline
PDH	&	1	&	HGHT	&	2	&	1.66E-02 $\pm$ 8.65E-04	&	-0.09 $\pm$ 0.26	\\
PDH	&	2	&	HGHT	&	2	&	1.78E-02 $\pm$ 3.92E-04	&	0.05 $\pm$ 0.12	\\
PDH	&	3	&	HGHT	&	1	&	1.82E-02 $\pm$ 3.99E-04	&	-0.42 $\pm$ 0.12	\\
PDH	&	4	&	HGHT	&	2	&	1.80E-02 $\pm$ 2.71E-04	&	0.01 $\pm$ 0.08	\\
PDH	&	5	&	HGHT	&	2	&	1.74E-02 $\pm$ 2.61E-04	&	0.04 $\pm$ 0.08	\\
PDH	&	6	&	HGHT	&	2	&	1.75E-02 $\pm$ 3.23E-04	&	0.05 $\pm$ 0.10	\\
PDH	&	7	&	HGHT	&	2	&	2.37E-02 $\pm$ 6.07E-04	&	-0.50 $\pm$ 0.18	\\
PDH	&	8	&	HGHT	&	2	&	2.97E-02 $\pm$ 3.17E-04	&	-0.17 $\pm$ 0.09	\\\hline\hline
HIDH	&	2	&	LGT	&	1	&	4.57E-01 $\pm$ 2.31E-02	&	7.07 $\pm$ 0.45	\\
HIDH	&	2	&	LGT	&	2	&	4.93E-01 $\pm$ 2.19E-02	&	7.40 $\pm$ 0.43	\\\hline

\label{table:CalFactors}
\end{tabular}
\end{table*}

These calibration coefficients are very different than the values published in \cite{radem_paper}. In fact, the slope obtained in this work is on average larger for the PDH HGHT triggers by a factor of 1.91, and smaller for the PDH HGLT, EDH HGLT, and HGHT triggers by a factor of 0.33, 0.14, and 0.96, respectively. Moreover, the values shown in \cite{radem_paper} did not consider the ASIC noise and offset. Although it is difficult to identify a single cause for these differences, variations in electrical and temperature conditions between the spacecraft and ground testing can affect the noise and gain of the detectors and ASICs. Additionally, the ground calibration in \cite{radem_paper} was performed using only two proton energies which can lead to large uncertainties in the calibration curves. 

\section{Flux reconstruction}\label{sec:reconstruction}
\begin{figure*}[h!]
\centering
\includegraphics[width=0.99\linewidth]{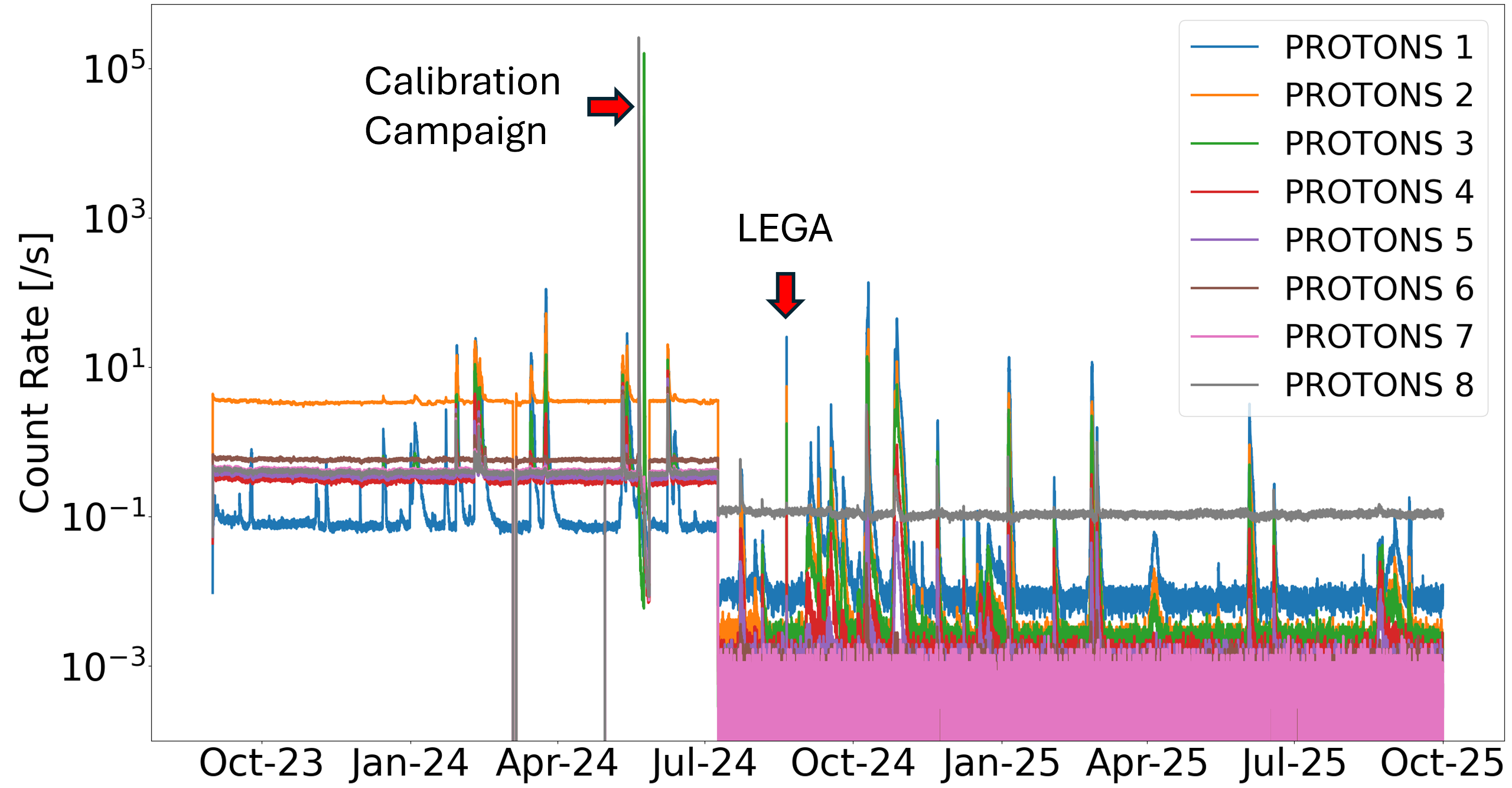}
     \caption{RADEM count rates of all PROTONS detection bins from August 30, 2023, until September 30, 2025. Each color represents a different detection bin. Most increases in count rates are cause by SEP events. The calibration scan executed between the 20th and 27th of May 2024, and the Earth gravity assist which took place on August 20 2024, can also be seen as increases in the count rates of most detection bins. The abrupt change on July 9th was due to a reconfiguration of the detection bins.}
     \label{fig:allprotons}
\end{figure*}

RADEM began acquiring science data on 30 August 2023. Figure~\ref{fig:allprotons} shows the count rates of the PROTONS detection bins up to October 2025. As seen in the figure, RADEM has already observed tens of SEPs as well as the Van Allen belts during the JUICE LEGA on August 20, 2024. The change in count rates (background) on July 9th, 2024, is due to a large reconfiguration (see Sect~\ref{sec:bins}) of the trigger thresholds and coincidence schemes of the detection bins. The exact threshold values and coincidence settings are shown in Appendix~\ref{app:config1}~-~\ref{app:config2} for both Configuration~1 and Configuration~2 respectively. 

\subsection{Science response functions}\label{sec:configs}
\begin{figure*}[]
\centering
\includegraphics[width=0.8\linewidth]{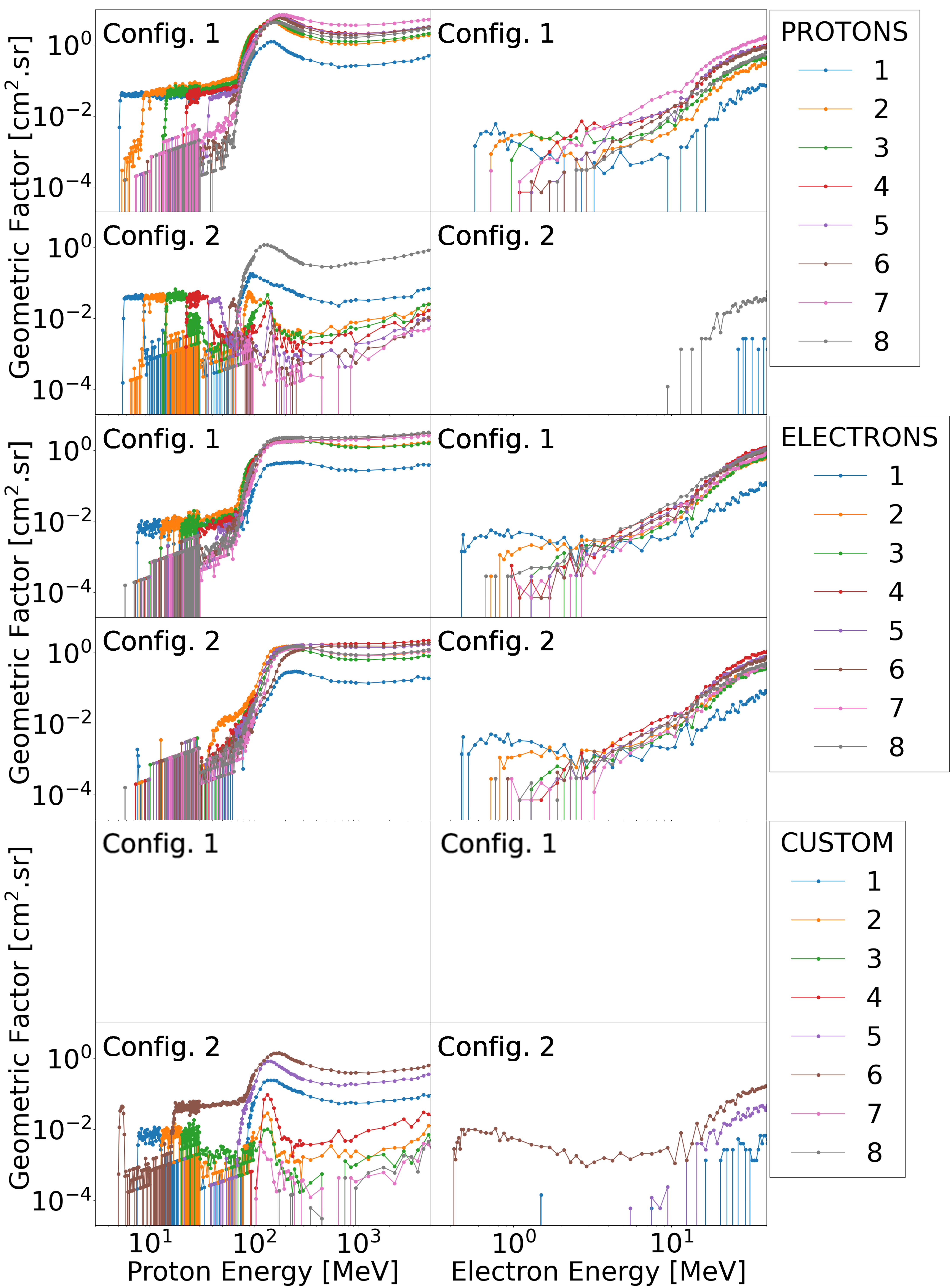}
     \caption{Response functions of the PROTONS (top four panels), ELECTRONS (middle four panels), and CUSTOM (bottom four panels) detection bins to protons (left column) and electrons (right column). Configuration~1 response functions are shown on top panels of each quadrant while Configuration~2 are shown in the bottom panels of each quadrant. The CUSTOM detection bins were not used in Configuration~1.}
     \label{fig:Configs}
\end{figure*}

Proton and electron response functions of the PROTONS, ELECTRONS, and the first five (corresponding to the EDH) CUSTOM detection bins are shown in Fig.~\ref{fig:Configs} for Configurations~1~and~2. They were calculated with the same methods described in Sect.~\ref{sec:calib} implementing the thresholds and coincidence logic of each detection bin.

Configuration~1 was focused on the detection of GCR protons \citep{radem_paper}. It was based on ground calibration curves and operated from August 30, 2023, to July 9, 2024. Each detector was configured in single coincidence mode, counting particles above the HGLT, which was set as low as possible above noise, and below the HGHT. As a result, all bins were sensitive to broad energy ranges of both protons and electrons, providing limited energy resolution and particle separation. Nevertheless, in \cite{LauraRADEM}, the proton flux at four energies 6.9, 13.3, 21.6, and 31.2~MeV were computed by combining detection bins and successfully cross-calibrated with STEREO-A's HET. A deviation less than 25\%, between RADEM and STEREO-A measurements was found. For this reason, in this paper we focus on Configuration 2 only. 

Configuration 2 has been in operation since July 9, 2024. It was defined to measure electron and proton spectra separately based on preliminary results from the GCR calibration. The main objective was to observe the Van Allen belts during the JUICE LEGA, and SEPs. In this configuration, PDH and EDH thresholds were selected to separate protons and electrons. The PDH detection bins were configured to use multiple detectors in coincidence and anti-coincidence, allowing for proton energy determination. The HGLT triggers were set to reject electron signals while the HGHT were selected to exclude heavy ions.
The EDH detection bins were configured to work in single coincidence mode, with all HGLT and HGHT triggers optimized to measure electrons - the HGLT was set as low as possible, and the HGHT was selected to exclude proton signals. Additionally, five EDH detection bins, defined as CUSTOM 1-5 in the RADEM raw data, were adapted to measure protons in coincidence mode. These detection bins are intended to inter-calibrate the PDH and the EDH detectors, and improve the proton energy resolution.
Although the LEGA measurements clearly demonstrated a distinction between electrons and protons, this analysis will not be addressed here, as a special issue is in preparation. During SEPs, electron measurements can be shadowed by proton contamination. In the case of RADEM, the ELECTRON detection bins are relatively sensitive to protons with energies above $\sim40-70$~MeV as depicted in the middle four panel of Fig.~\ref{fig:Configs}. For this reason, we will discuss electron measurements elsewhere and focus exclusively on the reconstruction of proton fluxes in this work.

\subsection{Bow-tie}
To provide physical meaning to RADEM's measurements, the particle flux must be derived from the observed count rates. In this work we used the bow-tie method, introduced by Van Allen~\citep{VanAllen1948BowTie,Pioneer11_paper}, to obtain energy dependent fluxes from the measured count rates. This method assumes a family of model energy spectra of the form:

\begin{equation}
    J(E) = J_0 E^{-\gamma}
    \label{eq:power_function}
\end{equation}

where $J(E)$ is the differential particle flux, $\gamma$ is the spectral index, and $J_0$ is a normalization factor. The first step of the method consists of defining a range of $\gamma$ values that are consistent with the radiation environment encountered by the instrument and convolving the model spectrum with the detector’s response function $RF(E)$, for each value of $\gamma$, as per Eq.~\ref{eq:CountRates}.

The next step of the method depends on the shape of the detection bin, which can be either integral or differential. Integral detection bins are sensitive to particles above a certain energy, whereas differential detection bins are sensitive to particles within an energy interval. The bow-tie curves are defined based on the expected count rate values as follows (differential Eq.\ref{eq:Bow_Tie_Diff}, integral Eq.\ref{eq:Bow_Tie_Int}):

\begin{equation}
    GdE(E) = \frac{\int_{0}^{\infty} RF(E') \cdot J(E') \, dE'}{J(E)}
    \label{eq:Bow_Tie_Diff}
\end{equation}

\begin{equation}
    G_I(E) = \frac{\int_{0}^{\infty} RF(E') \cdot J(E') \, dE}{\int_{E}^{\infty} J(E'')\, dE''} 
    \label{eq:Bow_Tie_Int}
\end{equation}

Where $GdE(E)$ and $G_I(E)$ are the energy-dependent geometric factors for differential and integral detection bins, respectively.
These curves are computed for the selected range of $\gamma$ values, generating a set of curves that forms the bow-tie. The point of minimal variance in this distribution (knot) defines the effective energy $E_\text{eff}$ and the corresponding geometric factor $G_\text{eff}$ of the detection bin.

The particle flux at $E_\text{eff}$ is then estimated by dividing the measured count rate $C_\text{rate}$ of each detection bin by the product of the effective geometric factor $G_\text{eff}$ and the integration time $\Delta t$:

\begin{equation}
    J(E_\text{eff}) = \frac{C_\text{rate}}{G_\text{eff}.\Delta E} 
\label{eq:fRecon}
\end{equation}

In this work, we selected a range of $\gamma$ between 2 and 5. This interval covers a wide variety of events while avoiding the most extreme cases where the bow-tie may face severe complications, either due to high energy proton contamination (small $\gamma$), or to insufficient counts in the response function range (large $\gamma$).

The results of the bow-tie method for protons in Configuration~1 and Configuration~2 can be found in Table~\ref{table:RADEMchannelsC1} and Table~\ref{table:RADEMchannelsC2} respectively. Configuration~1 has been discussed at length in \cite{LauraRADEM} and is shown here only for completeness. It is worth noting that in Configuration~1 different detection bins had to be combined to obtain a differential proton flux while in Configuration~2 each detection bin is used to derive the differential flux at a given effective energy. In the PROTONS detection bins of Configuration 2, the effective energy increases quasi-logarithmically as designed \citep{radem_paper}. The effective geometric factor, $G_\text{eff}$, varies by several orders of magnitude across detection bins due to a combination of energy range and detection efficiency. PROTONS~8 has a particularly high value, mostly due to its large energy range as it can be seen in Fig.~\ref{fig:Configs}. PROTONS~7 has a lower $G_\text{eff}$ than PROTONS~1, which, given its sensitivity to high energy protons, means it will always have the least counts of all the detection bins. 

\begin{table*}
\centering
\begin{threeparttable}
\caption{RADEM bow-tie results for protons in Configuration~1 including the list of linearly combined RADEM/PDH detection bins with their respective proton energy ranges and effective energy adapted from \cite{LauraRADEM}.}
\label{table:RADEMchannelsC1}

\begin{tabular}{cccccc}
\hline
Detection Bin Combination & \shortstack{Energy Range\\(MeV)} & 
\shortstack{Effective Energy\\(MeV)} & \shortstack{GdE\\(cm$^2\cdot$sr$\cdot$MeV)} & $\delta^-_G$ (\%) & $\delta^+_G$ (\%) \\
\hline\hline
(1) & (2) & (3) & (4) & (5) & (6) \\
\hline
1x[PROTONS 1] -- 1x[PROTONS 3] & 5.35--14.4 & 6.9  & 0.214 & -1.22 & 3.00 \\
1x[PROTONS 2] -- 1x[PROTONS 4] & 8.75--22.8 & 13.3 & 0.837 & -9.31 & 24.6 \\
1x[PROTONS 3] -- 1x[PROTONS 5] & 14.5--37.4 & 21.6 & 1.22  & -6.53 & 17.93 \\
1x[PROTONS 4] -- 1x[PROTONS 5] & 22.8--36.6 & 31.2 & 0.844 & -2.91 & 8.58 \\
\hline
\end{tabular}
\begin{tablenotes}
\footnotesize
\item[] \item[]\textbf{Notes.} Col. 1: Linear combination of proton detection bins used to obtain a differential detection bin.  
Col. 2: Proton energy range of the detection bin.  
Col. 3: Proton effective energy obtained with the bow-tie method.  
Col. 4: Mean value of the geometric factor distribution calculated with the bow-tie method.  
Cols. 5 and 6: 5th and 95th percentile of the geometric factor distribution subtracted from the mean value in percentage calculated with the bow-tie method.
\end{tablenotes}

\end{threeparttable}
\end{table*}

\begin{table*}
\centering
\begin{threeparttable}
\caption{RADEM bow-tie results for protons in Configuration 2 including energy ranges, effective energies and geometric factors.}
\label{table:RADEMchannelsC2}

\begin{tabular}{cccccc}
\hline
Detection Bin Combination & \shortstack{Energy Range\\(MeV)} & 
\shortstack{Effective Energy\\(MeV)} & \shortstack{GdE\\(cm$^2\cdot$sr$\cdot$MeV)} & $\delta^-_G$ (\%) & $\delta^+_G$ (\%) \\
\hline\hline
(1)&(2)&(3)&(4)&(5)&(6)\\
\hline
PROTONS 1 & 5.60 - 8.70 & 7.40 & 0.166 & 0.154 & 0.204 \\ 
PROTONS 2 & 8.80 - 14.3 & 11.1 & 0.233 & 0.226 & 0.252 \\
PROTONS 3 & 14.6 - 22.8 & 18.2 & 0.412 & 0.400 & 0.444 \\
PROTONS 4 & 22.8 - 36.5 & 28.9 & 0.601 & 0.587 & 0.639 \\ 
PROTONS 5 & 37.5 - 49.5 & 43.5 & 0.474 & 0.466 & 0.502 \\
PROTONS 6 & 58.5 - 67.5 & 66.5 & 0.352 & 0.352 & 0.380 \\
PROTONS 7 & 69.0 - 77.5 & 76.5 & 0.165 & 0.162 & 0.181 \\
PROTONS 8 & $\sim$70.0 - inf & 125 & 106 & 102 & 121\\ 
\hline
CUSTOM 1 & 7.80 - 12.8 & 14.8 & 0.215 & 0.152 & 0.390 \\
CUSTOM 2 & 12.9 - 20.2 & 17.2 & 0.078 & 0.074 & 0.092 \\
CUSTOM 3 & 20.3 - 30 & 27.4 &   0.115 & 0.110 & 0.130 \\
CUSTOM 4 & $\sim$70.0 - inf & 145 & 3.95 & 3.81 & 4.20 \\
CUSTOM 5 & $\sim$100 - inf & 135 & 71.5 & 66.7 & 84.6 \\ 
\hline
\end{tabular}

\begin{tablenotes}
\item[] \item[]\footnotesize
\textbf{Notes.} Col. 1: RADEM raw detection bin. Col. 2-6: Same as in Table~\ref{table:RADEMchannelsC1}.
\end{tablenotes}

\end{threeparttable}
\end{table*}

\subsection{Validation}
To validate both the calibration and flux reconstruction, we defined a series of tests using SEPs. Only protons were considered, as the ELECTRONS detection bins are contaminated by high-energy protons and therefore less reliable. For this purpose, we inter-calibrated the PDH (PROTONS) with the EDH (CUSTOM - EDH detection bins only), and compared RADEM proton fluxes with those from other missions.

\subsubsection{Intercalibration}
For the inter-calibration, we used the entire SEP dataset of Configuration 2, with the exception of the LEGA and the Venus gravity assist. For each point in time, the proton spectrum was considered to follow either a power-law (Eq. \ref{eq:power_function}) or a rolling power-law (Eq. \ref{eq:rol_power_function}). To make sure the fitting energy range fully covers the first three CUSTOM detection bins, we only considered periods when the count rate in the PROTONS~5 detection bin was 3 standard deviations ($\sigma$) above the background.

\begin{figure}[ht!]
\centering
\includegraphics[width=0.99\linewidth]{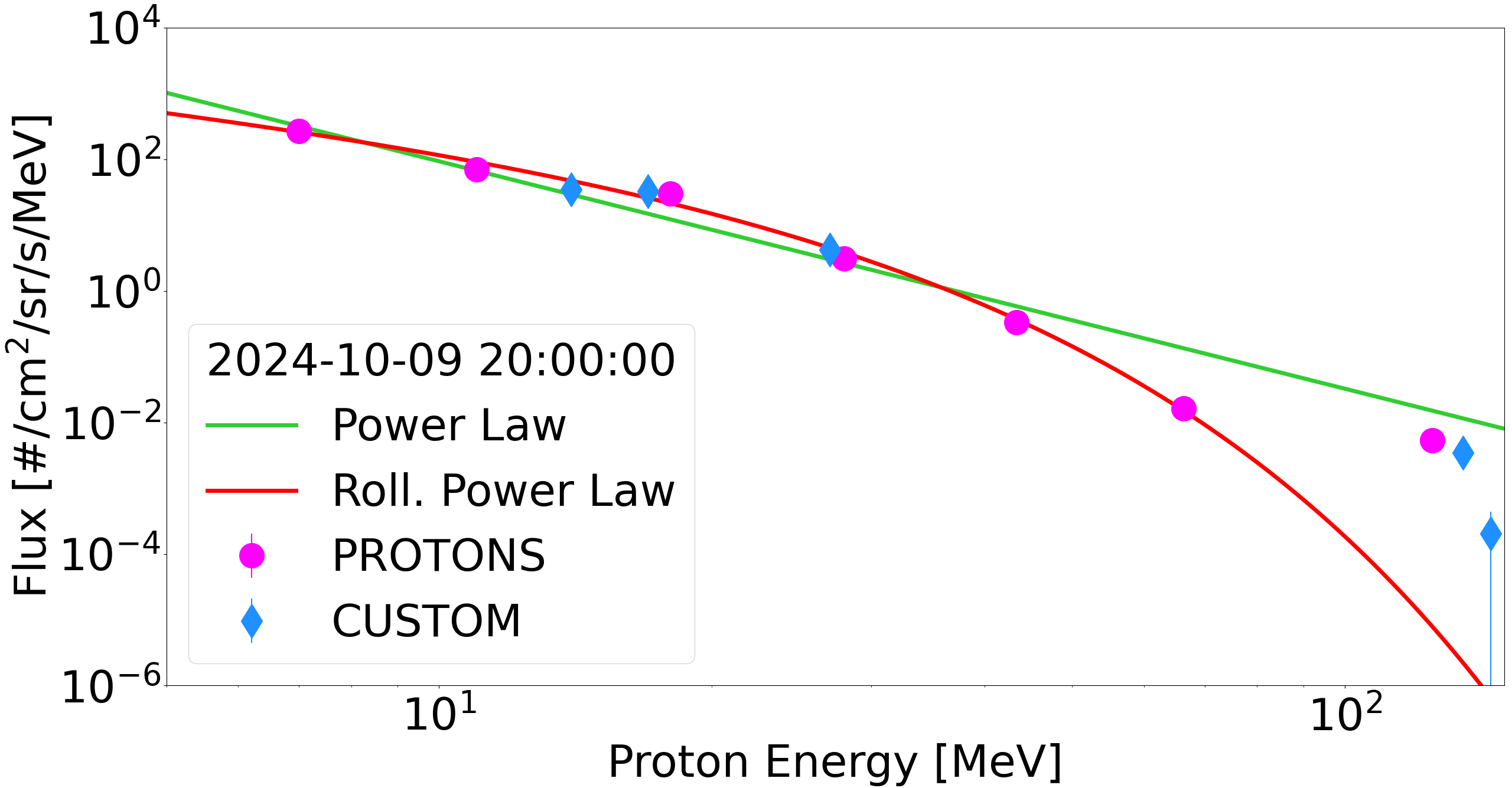}
     \caption{Proton flux as a function of energy acquired by RADEM on August 9th, 2024, between 20:00 and 21:00 UT. Magenta and blue points show the fluxes obtained with the PROTONS 1-7 and CUSTOM 1-3 detection bins, respectively. The result of a rolling power-law fit using PROTONS 1-6 detection bins is shown in red, while the best fit to a power-law using PROTONS~1-8 detection bins is shown in light green. A better fit is obtained with the rolling power-law which has been attributed to a larger effect of the spacecraft modeling in the bottom sensors.}
     \label{fig:InterCalibrationSpectralFit}
\end{figure}

As an example, Fig.~\ref{fig:InterCalibrationSpectralFit} shows the accumulated proton spectra acquired on October 9, 2024, between 20:00:00 and 21:00:00 UT. A rolling power-law fit was applied to the first six PROTONS detection bins. PROTONS~7 was excluded due to its negligible count rate and PROTONS~8 was removed from the analysis because it led to a very large deviation of several detection bins with relation to the fit. The fitted spectrum is in good agreement with the proton flux computed with the first three CUSTOM detection bins (within $\sim$33\%). This provides strong evidence that the proton spectrum is well reconstructed, since the CUSTOM detection bins are defined with the EDH while the PROTONS detection bins are defined with the PDH — two completely independent systems, both in terms of sensors and front-end electronics. 

PROTONS~8 and CUSTOM~4-5 detection bins overestimate the flux obtained from the rolling power-law fit at the same effective energy by more than two orders of magnitude. As shown in Fig.~\ref{fig:InterCalibrationSpectralFit}, when fitting a simple power-law to all PROTON detection bins (with the exception of PROTONS~7 which had no counts), the PROTONS~3 and CUSTOM~2 detection bins overestimate the flux by a factor of 2-3, while the PROTONS~6, PROTONS~8 and CUSTOM~4-5 detections bins underestimate the flux by at least a factor of three. These differences are likely due to two reasons. First, PROTONS~8 and CUSTOM~4-5 detection bins have a coincidence pattern that counts particles coming from the bottom of the PDH and EDH, respectively. During cruise, RADEM FOV is pointing away from the Sun which means that these detection bins primarily measure particles from the Sun-JUICE direction. When the particle flux is not isotropic, their flux is not comparable to the rest of the detection bins. Second, RADEM Geant4 model does not include spacecraft material beneath the instrument, as shown in Fig.~\ref{fig:GEANT4}. Even though this model was provided by the JUICE manufacturer, it appears unrealistic given that the JUICE high-gain antenna is on the opposite panel of the spacecraft. Underestimating the shielding results in a larger geometric factor and therefore, a smaller flux. However, if we consider the calibration, underestimating the shielding changes the mapping between flight energy threshold and simulation energy threshold. This means that a higher simulation threshold matches the flight threshold since it has more counts than what it should. As shown in Fig.~\ref{fig:ThRFsP}, a higher threshold results in a smaller geometric factor which in turn leads to a higher flux. Spacecraft modeling is expected to have a stronger impact on the bottom detectors of the PDH. It was not observed in the calibration coefficients of the EDH and HIDH, which can be attributed to their smaller stack dimensions that place them farther away from the bottom panel (see Fig.~\ref{fig:detheads}).

PROTONS~8 and CUSTOM~4-5 detection bins overestimate the flux obtained from the rolling power-law fit at the same effective energy by more than two orders of magnitude. As shown in Fig.~\ref{fig:InterCalibrationSpectralFit}, when fitting a simple power-law to all PROTON detection bins (except PROTONS~7, which had no counts), the PROTONS~3 and CUSTOM~2 detection bins overestimate the flux by a factor of 2–3, while PROTONS~6, PROTONS~8, and CUSTOM~4-5 detection bins underestimate the flux by at least a factor of three. These differences are likely due to two reasons.  

First, PROTONS~8 and CUSTOM~4-5 detection bins have a coincidence pattern that counts particles coming from the bottom of the PDH and EDH, respectively. During cruise, RADEM FOV points away from the Sun, so these detection bins primarily measure particles from the Sun–JUICE direction. When the particle flux is anisotropic, their flux is not directly comparable to the other detection bins.  

Second, the RADEM Geant4 model does not include spacecraft material beneath the instrument, as shown in Fig.~\ref{fig:GEANT4}. Although this model was provided by the JUICE manufacturer, it appears unrealistic given that the JUICE high-gain antenna is on the opposite panel. Underestimating the shielding affects the geometric factor in two ways. Less shielding allows more particles to reach the detector, which increases the simulated count rate and would reduce the inferred flux. However, in the calibration procedure, detector thresholds (DAC units) are mapped to simulation energies. With underestimated shielding, a given flight threshold corresponds to a higher simulated energy, which reduces the integrated flux in the simulation and therefore leads to a larger derived geometric factor and a higher calibrated flux. This effect is expected to have the strongest impact on the bottom detectors of the PDH and EDH. It was not observed in the calibration coefficients of the HIDH, which can be attributed to their smaller stack dimensions and greater distance from the bottom panel (see Fig.~\ref{fig:detheads}).

\begin{figure}[h!]
\centering
\includegraphics[width=0.99\linewidth]{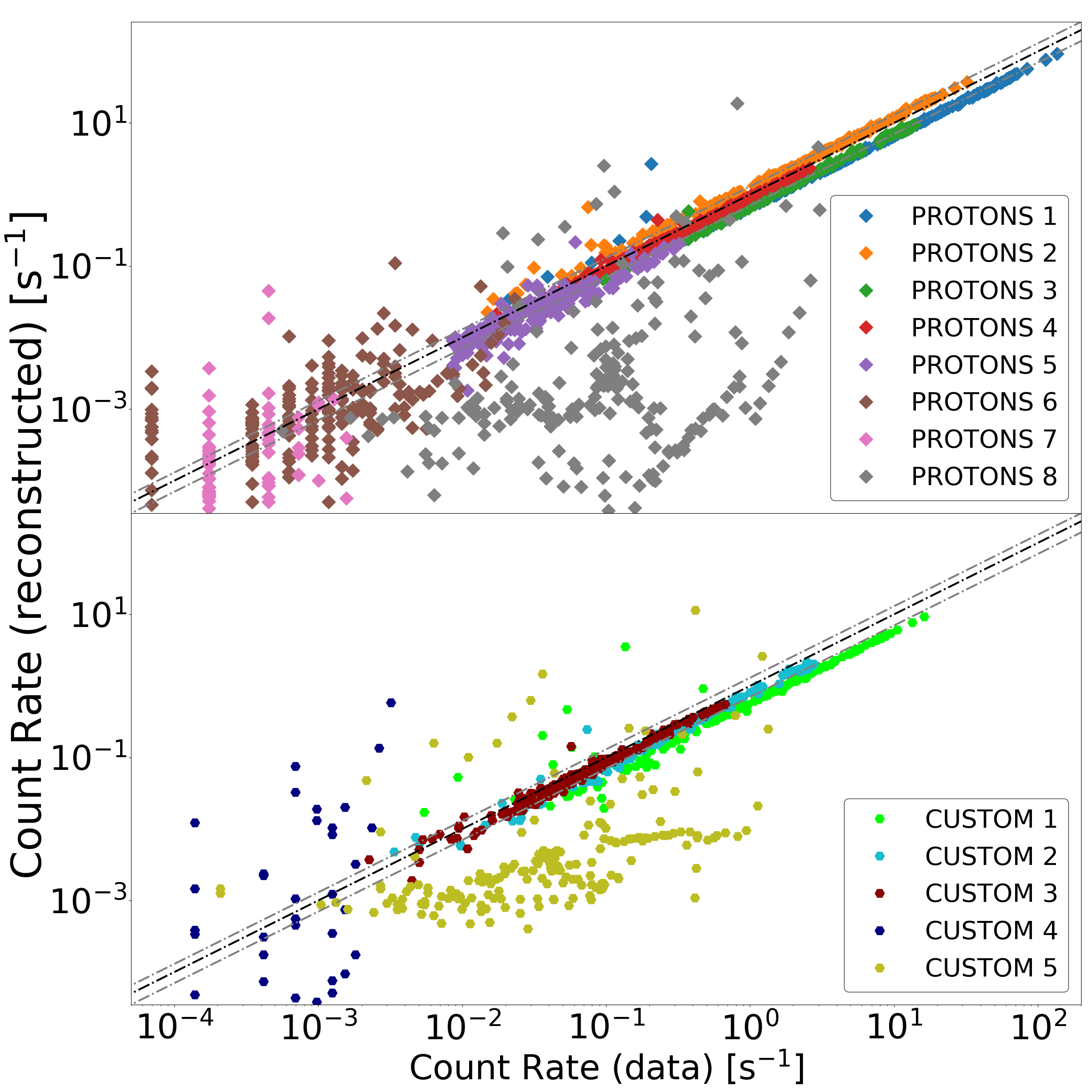}
     \caption{Comparison between the flight rates observed in the PROTONS (top) and CUSTOM (bottom) detection bins and the rates reconstructed using a rolling power-law fit for each point in time with at least five detection bins above the background. Each color represents a different detection bin. The black line represents a perfect reconstruction and the two lines in gray represent a 30\% deviation.}
     \label{fig:InterCalibrationReconAll}
\end{figure}

To further evaluate the quality of the bow-tie method, we reconstructed the flight rate of all PROTONS and CUSTOM detection bins with Eq.~\ref{eq:fRecon} using the fitted spectra. For this purpose, only time periods of 1-hour when the first five PROTONS detection bins had count rates 3$\sigma$ above the background were selected. 
Comparison between the reconstructed count rates with the flight count rates for each period are shown in Fig.~\ref{fig:InterCalibrationReconAll}. The reconstructed rates of PROTONS~1-6 and CUSTOM~1-3 detection bins are on average within $\sim$30-50\% of the flight rates, providing strong evidence that our method is effective. Since the variance for these detections bins is small, we attribute the differences between the flight and reconstructed count rates to the bow-tie analysis and the rolling power-law fit. 
On the other hand, the reconstructed rates of PROTONS~7-8 and CUSTOM~4-5 detection bins did not match the flight count rates. For PROTONS~7 and CUSTOM~5 detection bins, this can be easily explained by the lack of statistically relevant counts above background. For PROTONS~8 and CUSTOM~4 detection bins, we associate this discrepancy to the incomplete spacecraft modeling as mentioned before. Nevertheless, these results strongly support the methods employed this study to calibrate RADEM and to calculate the proton flux.

\subsubsection{Comparison with SOHO/ERNE}
Even though the PDH and EDH are completely detached systems, it remains essential to compare RADEM measurements with independent observations. Since instruments on JUICE such as the Particle Environment Package \citep{Barabash2016_PEP_JUICE} are off most of the cruise to Jupiter, we rely on particle detectors from other missions for comparison.

\begin{figure}[!ht]
\centering
\includegraphics[width=0.99\linewidth]{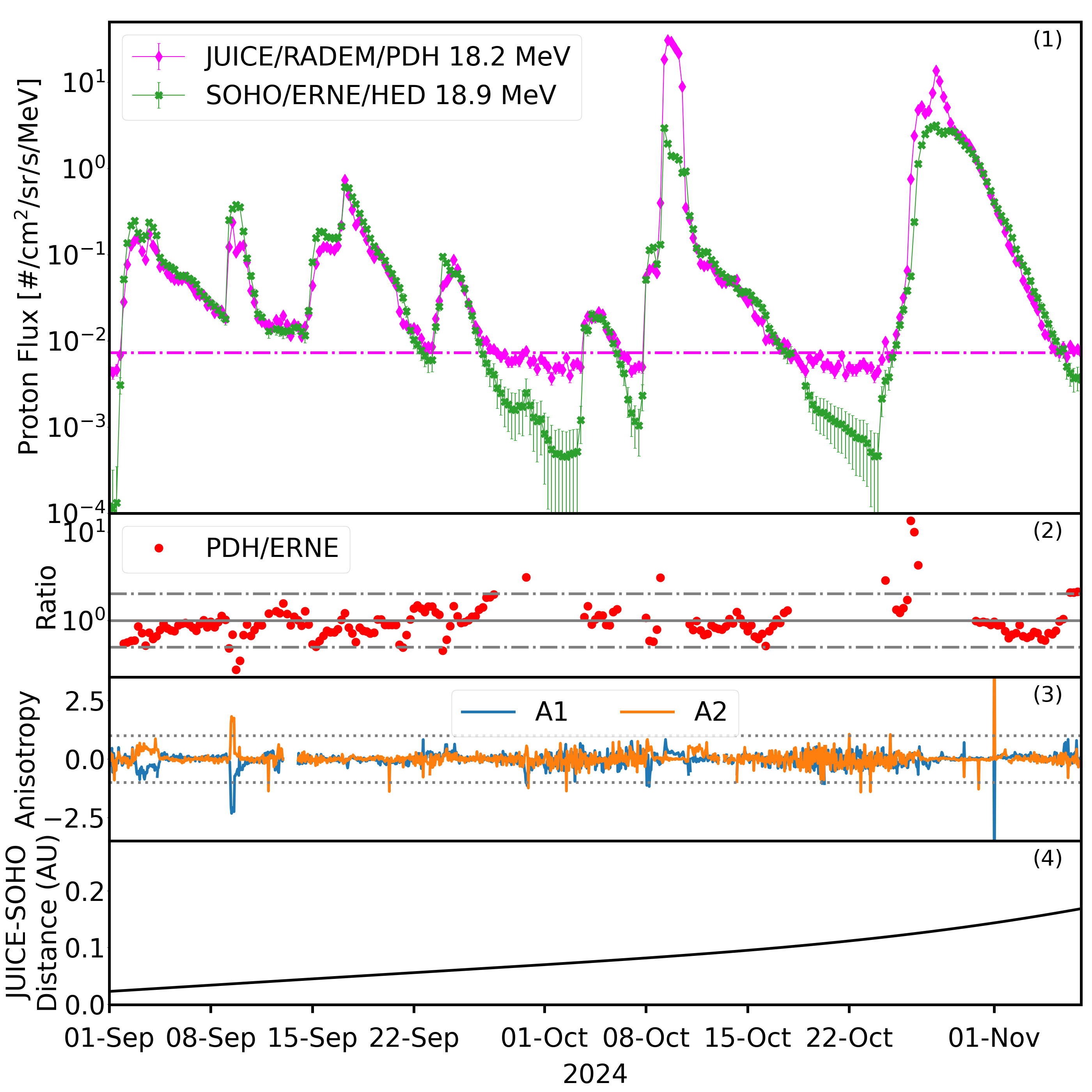}
     \caption{Comparison of the proton flux measured by RADEM and ERNE from the 1st of September to the 6th of November, 2025.From top to bottom: (1) proton flux measured by RADEM/PDH (pink) and ERNE/HED (green); (2) ratio between RADEM/PDH and ERNE/HED when RADEM/PDH flux is three standard deviations above background (dot-dashed pink line in the top panel), and A1 and A2 are between -1 and 1; (3) first and second-order anystropies (A1 and A2) calculated using the Wind spacecraft 3D Plasma Analyzed; and (4) JUICE-SOHO distance. Further details are given in the main text.}
     \label{fig:ComparisonPDH-HED}
\end{figure}

In September and October 2024, RADEM observed multiple SEP events as shown in Fig.~\ref{fig:ComparisonPDH-HED}. These events took place shortly after the LEGA, when JUICE was still relatively close to Earth. ERNE aboard SOHO orbiting the Lagrange point 1 also measured these events (see Fig.~\ref{fig:ComparisonPDH-HED}) with its the High Energy Detector \citep [HED;][]{ERNE_paper}. During this period, the distance between JUICE and SOHO varied from $\sim$0.02~au $\sim$to 0.17~au as depicted in the bottom panel of Fig.~\ref{fig:ComparisonPDH-HED}. For these reasons, we took this opportunity to compare the proton flux recorded by RADEM with those obtained by ERNE. It is worth noting that small distances between spacecrafts does not necessarily imply that they sample the same particle populations \citep{Khoo2024BERMvsPSP}. In fact, during the JUICE cruise phase, RADEM's detectors are pointing away from the Sun while ERNE measures particles coming from the Sun. This is due to JUICE operational constraints and can have a great effect on the measured flux if the pitch angle distribution is not isotropic \citep{LauraRADEM}.

To identify periods when the proton flux environment was similar across the regions surrounding the analyzed instruments, we computed the anisotropy from Wind spacecraft's 3D Plasma Analyzer measurements \citep{wind3dp, ogilvie1997wind}. An isotropic flux may indicate the presence of a reservoir effect, which typically occurs during the decay phase of an SEP event. This effect is characterized by spatial invariance in the spectra and, consequently, by isotropic particle fluxes \citep{Reames_1997,Wang_2021,2010Lario}. We calculated the first-order and second-order anisotropies, A1 and A2 respectively, with the Energetic Solar Eruptions: Data and Analysis Tools \citep[SOLER;][]{gieseler2025}, based on data from the eight viewing directions of Wind. Legendre polynomials were fitted to the data, whose results are shown in Fig.~\ref{fig:ComparisonPDH-HED}. As it can be seen in the third panel of Fig.~\ref{fig:ComparisonPDH-HED}, A1 and A2 were between -1 and 1 most of the time which typically indicates an isotropic flux. This is unusual, especially during the onset of SEP events but suggests that the measurements of the two instruments can be compared. 

The ratio between the proton flux measured by RADEM/PDH and ERNE/HED for similar energies—18.2 MeV and 18.9 MeV, respectively—can be seen in the second panel from the top in Fig.~\ref{fig:ComparisonPDH-HED}. Two criteria were used to calculate this ratio. First, since the RADEM background is much higher than ERNE’s at these energies, we only considered points in time when the RADEM flux was three standard deviations above the background. Second, to ensure that the flux was isotropic, we excluded periods when A1 and A2 were above 1 or below –1. In general, the flux measured by the two instruments agrees within a factor of two. The only exceptions occur during the onsets of a few events, when the RADEM proton flux is very close to the background (within three standard deviations), and when ERNE/HED saturates.
The latter can be seen in the last two and more intense events. While ERNE/HED never registers a proton flux above $\sim3.18~\#/cm^2/sr/s/MeV$, RADEM/PDH measures proton flux up to an order of magnitude higher at a similar energy. We removed the data points when ERNE/HED saturated from our analysis. Notice that this does not exclude three points during the onset of the last event where a large mismatch can be seen. This mismatch seems to be related to particle propagation and not necessarily to ERNE/HER saturation.
With these considerations, we calculated a mean ratio of 1.11 for this pair of energies. For the other energy pairs (28.9/29.0 MeV and 43.45/45.6 MeV) where the RADEM proton flux is much closer to the background, we obtained ratios of 0.82 and 0.81, respectively. While these values should not be used for cross-calibration they provide confidence in the RADEM proton flux reconstruction. These ratios are likely slightly overestimated due to the onset of the last event.

\section{Summary and discussion}\label{sec:discussion}
RADEM is a facility instrument aboard the JUICE mission launched on April 14, 2023 \citep{JUICE_paper,radem_paper}. It was designed to measure energetic protons, electrons, and ions in interplanetary space and at Jupiter. RADEM has three silicon-stack telescopes, the EDH, the PDH, and the HIDH These detectors use trigger patterns to identify particle species and energies. RADEM also includes a DDH that was not analyzed in this paper.
Since August 30, 2023, RADEM has recorded more than 30 SEPs events. However, the flight data could not be reproduced using the pre-launch ground calibration.

For this reason, we designed a calibration procedure in interplanetary space using GCRs. The method consisted of scanning the detectors trigger thresholds and compare the change in count rate with Geant4 simulations. Unfortunately, during the calibration a campaign, an SEP event contaminated the data but we used close-by spacecraft (STEREO-A) measurements to minimize its impact. 

The calibration coefficients obtained were in disagreement with the ground-calibration which reflects limitations of the pre-launch calibration in reproducing the detector response under flight conditions. In particular, the ground calibration was based on a limited number of data points and did not fully account for electronic noise, ASIC processing effects, and offset terms present during in-flight operation \citep{radem_paper}. As a result, the ground calibration does not provide a complete representation of the instrument response in space, motivating the need for an in-flight calibration using GCRs. Additionally, spacecraft shielding was identified as a source of uncertainty, particularly affecting the calibration factors of the bottom-most detectors of the PDH and EDH.

To evaluate our results, we reconfigured RADEM and calculated its response functions based on the new calibration coefficients. We then compared RADEM SEP measurements acquired in September and October 2024 while JUICE was close to the Earth, to those of SOHO/ERNE orbiting the Lagragian point 1. RADEM proton flux measurements were in good agreement (within a factor of two) with ERNE even though the two were not necessarily sampling the same particle population. In fact, RADEM is susceptible to directional effects since its detector heads have narrow FOVs and are pointing away from the Sun during most of the JUICE cruise phase while ERNE measures particles coming from the Sun. 

The method described in this paper will be performed repeatedly to improve RADEM's calibration and monitor any possible degradation. Opportunities to cross-calibrate the instrument with others on the JUICE mission such as PEP \citep{Barabash2016_PEP_JUICE}, and in other spacecraft such as Solar Orbiter's Energetic Particle Detector \citep{Rodriguez-Pacheco2020_EPDSolarOrbiter}, Parker Solar Probe's Integrated Science Investigation of the Sun instrument \citep{McComas2016_ISIS_PSP}, and BepiColombo's Environment Radiation Monitor \citep{BERM}, will also be sought. Particular attention will be given to SEP events that occur where JUICE is close to other spacecraft and/or when the reservoir effect is present.

\section{Conclusions}\label{sec:conclusions}
In this work we successfully developed a calibration method based on GCRs and applied it to to the JUICE radiation monitor, RADEM. The results were validated by comparing JUICE/RADEM SEP measurements with those from SOHO/ERNE when the two spacecraft were relatively close. The derived fluxes agree within a factor of two, demonstrating that RADEM provides reliable energetic proton data. This means that even though RADEM is a facility instrument aboard a planetary mission, it is a valuable tool for heliophysics science, contributing to the characterization of SEPs and improving our understanding of the radiation environment along the JUICE trajectory.

\begin{acknowledgements}
    JUICE is a mission led by the European Space Agency (ESA), with major contributions from its Member States, the National Aeronautics and Space Administration (NASA), the Japan Aerospace Exploration Agency (JAXA), and the Israel Space Agency. RADEM data are available from the ESA Planetary Science Archive (PSA) at https://www.psa.esa.int/.
    
    The work of M. Pinto, F. Santos, A. Gomes, L. Arruda and P. Gonçalves was performed under an ESA contract: 4000137865/22/ES/JDExpert Suport to BERM $\&$ RADEM units. A. Gomes' work was funded was funded by the Portuguese Foundation for Science and Technology (FCT) through the research grant no. UI/BD/154742/2023, under the framework of the project 'Advanced Methods for Solar Energetic Particle Events Characterization in the Inner Solar System' 
    The authors would like to thank the JUICE team including Ry Evill and Angela Dietz for their help setting the calibration runs and Joana S. Oliveira and Fran Vallejo for their contributions in archiving the RADEM data.
\end{acknowledgements}


\bibliographystyle{bibtex/aa}
\bibliography{bibtex/biblio}

\begin{appendix}
\onecolumn

\section{Configuration 1}\label{app:config1}
\FloatBarrier   

\begin{table}[H]
\centering
\caption{Thresholds values for each detector trigger in Configuration 1.}
\label{table:RADEMthresholdConfig1}

\begin{tabular}{ccccc}
\hline
Detector & HGLT [DAC] & HGHT [DAC] & HGLT [MeV] & HGHT [MeV] \\
\hline
EDH 1 & 153 & 422 & 0.19 & 7.60 \\
EDH 2 & 153 & 449 & 0.16 & 7.15 \\
EDH 3 & 143 & 460 & 0.17 & 8.01 \\
EDH 4 & 153 & 351 & 0.06 & 5.69 \\
EDH 5 & 143 & 335 & 0.08 & 5.74 \\
EDH 6 & 143 & 406 & 0.05 & 6.84 \\
EDH 7 & 122 & 322 & 0.10 & 5.31 \\
EDH 8 & 112 & 389 & 0.09 & 7.04 \\ \hline \hline
PDH 1 & 378 & 817 & 0.52 & 13.1 \\
PDH 2 & 368 & 707 & 0.51 & 12.2 \\
PDH 3 & 347 & 666 & 0.47 & 11.4 \\
PDH 4 & 337 & 852 & 0.33 & 14.9 \\
PDH 5 & 347 & 977 & 0.34 & 16.5 \\ 
PDH 6 & 368 & 783 & 0.38 & 13.4 \\
PDH 7 & 265 & 411 & 0.34 & 9.01 \\
PDH 8 & 378 & 565 & 0.66 & 16.1 \\ \hline

\end{tabular}
\end{table}

\begin{table}[h!]
\centering
\caption{PDH and HIDH ASIC Coincidence 1 scheme (C: Coincidence, AC: Anti-Coincidence).}
\label{table:config1pdh}
\small
\setlength{\tabcolsep}{4pt}
\begin{tabular}{lcc*{16}{c}}
\hline
Detection Bin & \multicolumn{2}{c}{Low Gain} & \multicolumn{16}{c}{High Gain} \\
\cline{2-3}\cline{4-19}
 & HIDH D1 & HIDH D2 & \multicolumn{2}{c}{PDH D1} & \multicolumn{2}{c}{PDH D2} & \multicolumn{2}{c}{PDH D3} & \multicolumn{2}{c}{PDH D4} &
   \multicolumn{2}{c}{PDH D5} & \multicolumn{2}{c}{PDH D6} & \multicolumn{2}{c}{PDH D7} & \multicolumn{2}{c}{PDH D8} \\
\cline{2-3}\cline{4-5}\cline{6-7}\cline{8-9}\cline{10-11}\cline{12-13}\cline{14-15}\cline{16-17}\cline{18-19}
 & LGT1 & LGT2 & LT & HT & LT & HT & LT & HT & LT & HT & LT & HT & LT & HT & LT & HT & LT & HT \\
\hline
HEAVY\_IONS 1 & C & X & X & X & X & X & X & X & X & X & X & X & X & X & X & X & X & X \\
HEAVY\_IONS 2 & X & C & X & X & X & X & X & X & X & X & X & X & X & X & X & X & X & X \\\hline\hline
PROTONS 1     & X & X & C & AC & X & X & X & X & X & X & X & X & X & X & X & X & X & X \\
PROTONS 2     & X & X & X & X  & C & AC & X & X & X & X & X & X & X & X & X & X & X & X \\
PROTONS 3     & X & X & X & X  & X & X  & C & AC & X & X & X & X & X & X & X & X & X & X \\
PROTONS 4     & X & X & X & X  & X & X  & X & X  & C & AC & X & X & X & X & X & X & X & X \\
PROTONS 5     & X & X & X & X  & X & X  & X & X  & X & X  & C & AC & X & X & X & X & X & X \\
PROTONS 6     & X & X & X & X  & X & X  & X & X  & X & X  & X & X  & C & AC & X & X & X & X \\
PROTONS 7     & X & X & X & X  & X & X  & X & X  & X & X  & X & X  & X & X  & C & AC & X & X \\
PROTONS 8     & X & X & X & X  & X & X  & X & X  & X & X  & X & X  & X & X  & X & X  & C & AC \\
\hline
\end{tabular}
\end{table}

\begin{table}[h!]
\centering
\caption{EDH ASIC Coincidence 1 scheme (C: Coincidence, AC: Anti-Coincidence).}
\label{table:config1edh}
\small
\setlength{\tabcolsep}{4pt}
\begin{tabular}{lcc*{16}{c}}
\hline
Detection Bin & \multicolumn{2}{c}{Low Gain} & \multicolumn{16}{c}{High Gain} \\
\cline{2-3}\cline{4-19}
 & NA & NA & \multicolumn{2}{c}{EDH D1} & \multicolumn{2}{c}{EDH D2} & \multicolumn{2}{c}{EDH D3} & \multicolumn{2}{c}{EDH D4} &
   \multicolumn{2}{c}{EDH D5} & \multicolumn{2}{c}{EDH D6} & \multicolumn{2}{c}{EDH D7} & \multicolumn{2}{c}{EDH D8} \\
\cline{2-3}\cline{4-5}\cline{6-7}\cline{8-9}\cline{10-11}\cline{12-13}\cline{14-15}\cline{16-17}\cline{18-19}
 & LGT1 & LGT2 & LT & HT & LT & HT & LT & HT & LT & HT & LT & HT & LT & HT & LT & HT & LT & HT \\
\hline
ELECTRONS 1 & X & X & C  & AC & X & X & X & X & X & X & X & X & X & X & X & X & X & X \\
ELECTRONS 2 & X & X & X  & X  & C & AC & X & X & X & X & X & X & X & X & X & X & X & X \\
ELECTRONS 3 & X & X & X  & X  & X & X  & C & AC & X & X & X & X & X & X & X & X & X & X \\
ELECTRONS 4 & X & X & X  & X  & X & X  & X & X  & C & AC & X & X & X & X & X & X & X & X \\
ELECTRONS 5 & X & X & X  & X  & X & X  & X & X  & X & X  & C & AC & X & X & X & X & X & X \\
ELECTRONS 6 & X & X & X  & X  & X & X  & X & X  & X & X  & X & X  & C & AC & X & X & X & X \\
ELECTRONS 7 & X & X & X  & X  & X & X  & X & X  & X & X  & X & X  & X & X  & C & AC & X & X \\
ELECTRONS 8 & X & X & X  & X  & X & X  & X & X  & X & X  & X & X  & X & X  & X & X  & C & AC \\\hline\hline
CUSTOM 1    & X & X & X  & X  & X & X  & X & X  & X & X  & X & X  & X & X  & X & X  & X & X \\
CUSTOM 2    & X & X & X  & X  & X & X  & X & X  & X & X  & X & X  & X & X  & X & X  & X & X \\
CUSTOM 3    & X & X & X  & X  & X & X  & X & X  & X & X  & X & X  & X & X  & X & X  & X & X \\
CUSTOM 4    & X & X & X  & X  & X & X  & X & X  & X & X  & X & X  & X & X  & X & X  & X & X \\
CUSTOM 5    & X & X & X  & X  & X & X  & X & X  & X & X  & X & X  & X & X  & X & X  & X & X \\
\hline
\end{tabular}
\end{table}

\newpage
\section{Configuration 2}\label{app:config2}
\FloatBarrier   


\begin{table}[h!]
\centering
\caption{Thresholds values for each detector trigger in Configuration 2.}
\label{table:RADEMthresholdConfig2}
\begin{tabular}{ccccc}
\hline
Detector & HGLT [DAC] & HGHT [DAC] & HGLT [MeV] & HGHT [MeV] \\
\hline
EDH 1 & 180 & 40 & 0.23 & 5.54 \\
EDH 2 & 180 & 40 & 0.21 & 0.94 \\
EDH 3 & 180 & 40 & 0.23 & 0.65 \\
EDH 4 & 180 & 40 & 0.11 & 0.72 \\
EDH 5 & 180 & 40 & 0.14 & 0.78 \\
EDH 6 & 180 & 40 & 0.11 & 0.44 \\
EDH 7 & 180 & 40 & 0.20 & 0.69 \\
EDH 8 & 180 & 40 & 0.21 & 0.73 \\ \hline \hline
PDH 1 & 280 & 120 & 0.37 & 1.90 \\
PDH 2 & 800 & 600 & 1.30 & 10.4 \\
PDH 3 & 800 & 600 & 1.27 & 10.2 \\
PDH 4 & 800 & 600 & 1.17 & 10.5 \\
PDH 5 & 800 & 600 & 1.15 & 10.2 \\ 
PDH 6 & 800 & 600 & 1.16 & 10.3 \\
PDH 7 & 800 & 600 & 1.41 & 13.4 \\
PDH 8 & 800 & 600 & 1.65 & 17.1 \\ \hline
\end{tabular}
\end{table}

\begin{table}[h!]
\centering
\caption{PDH and HIDH ASIC Coincidence 2 scheme (C: Coincidence, AC: Anti-Coincidence).}
\small
\label{table:PDHchannelMappingConfig2}
\setlength{\tabcolsep}{4pt}
\begin{tabular}{lcc*{16}{c}}
\hline
Detection Bin & \multicolumn{2}{c}{Low Gain} & \multicolumn{16}{c}{High Gain} \\
\cline{2-3}\cline{4-19}
 & HIDH D1 & HIDH D2 & \multicolumn{2}{c}{PDH D1} & \multicolumn{2}{c}{PDH D2} & \multicolumn{2}{c}{PDH D3} & \multicolumn{2}{c}{PDH D4} &
   \multicolumn{2}{c}{PDH D5} & \multicolumn{2}{c}{PDH D6} & \multicolumn{2}{c}{PDH D7} & \multicolumn{2}{c}{PDH D8} \\
\cline{2-3}\cline{4-5}\cline{6-7}\cline{8-9}\cline{10-11}\cline{12-13}\cline{14-15}\cline{16-17}\cline{18-19}
 & LGT1 & LGT2 & LT & HT & LT & HT & LT & HT & LT & HT & LT & HT & LT & HT & LT & HT & LT & HT \\
\hline
HEAVY\_IONS 1 & X & X & X & X & X & X & X & X & X & X & X & X & AC & X & X & C & X & C \\
HEAVY\_IONS 2 & X & C & X & X & X & X & X & X & X & X & X & X & X  & X & X & X & X & X \\ \hline\hline
PROTONS 1     & X & X & X & C & AC & X & X & X & X & X & X & X & X  & X & X & X & X & X \\
PROTONS 2     & X & X & X & C & C  & AC & AC & X & X & X & X & X & X & X & X & X & X & X \\
PROTONS 3     & X & X & C & X & X  & X  & C  & AC & AC & X & X & X & X & X & X & X & X & X \\
PROTONS 4     & X & X & C & X & X  & X  & X  & X  & C  & AC & AC & X & X & X & X & X & X & X \\
PROTONS 5     & X & X & C & X & X  & X  & X  & X  & X  & X  & C  & AC & AC & X & X & X & X & X \\
PROTONS 6     & X & X & C & X & X  & X  & X  & X  & X  & X  & X  & X  & C  & AC & AC & X & X & X \\
PROTONS 7     & X & X & C & X & X  & X  & X  & X  & X  & X  & X  & X  & X  & X  & C  & AC & AC & X \\
PROTONS 8     & X & X & X & X & X  & X  & X  & X  & X  & X  & X  & X  & X  & X  & AC & X & C  & AC \\
\hline
\end{tabular}
\end{table}

\begin{table}[h!]
\centering
\caption{EDH ASIC Coincidence 2 scheme (C: Coincidence, AC: Anti-Coincidence).}
\label{table:EDHchannelMappingConfig2}
\small
\setlength{\tabcolsep}{4pt}
\begin{tabular}{lcc*{16}{c}}
\hline
Detection Bin & \multicolumn{2}{c}{Low Gain} & \multicolumn{16}{c}{High Gain} \\
\cline{2-3}\cline{4-19}
 & NA & NA & \multicolumn{2}{c}{EDH D1} & \multicolumn{2}{c}{EDH D2} & \multicolumn{2}{c}{EDH D3} & \multicolumn{2}{c}{EDH D4} &
   \multicolumn{2}{c}{EDH D5} & \multicolumn{2}{c}{EDH D6} & \multicolumn{2}{c}{EDH D7} & \multicolumn{2}{c}{EDH D8} \\
\cline{2-3}\cline{4-5}\cline{6-7}\cline{8-9}\cline{10-11}\cline{12-13}\cline{14-15}\cline{16-17}\cline{18-19}
 & LGT1 & LGT2 & LT & HT & LT & HT & LT & HT & LT & HT & LT & HT & LT & HT & LT & HT & LT & HT \\
\hline
ELECTRONS 1   & X & X & C  & AC & X & X & X & X & X & X & X & X & X & X & X & X & X & X \\
ELECTRONS 2   & X & X & X  & X  & C & AC & X & X & X & X & X & X & X & X & X & X & X & X \\
ELECTRONS 3   & X & X & X  & X  & X & X  & C & AC & X & X & X & X & X & X & X & X & X & X \\
ELECTRONS 4   & X & X & X  & X  & X & X  & X & X  & C & AC & X & X & X & X & X & X & X & X \\
ELECTRONS 5   & X & X & X  & X  & X & X  & X & X  & X & X  & C & AC & X & X & X & X & X & X \\
ELECTRONS 6   & X & X & X  & X  & X & X  & X & X  & X & X  & X & X  & C & AC & X & X & X & X \\
ELECTRONS 7   & X & X & X  & X  & X & X  & X & X  & X & X  & X & X  & X & X  & C & AC & X & X \\
ELECTRONS 8   & X & X & X  & X  & X & X  & X & X  & X & X  & X & X  & X & X  & X & X  & C & AC \\ \hline\hline
CUSTOM 1       & X & X & X  & C  & AC & X  & X & X  & X & X  & X & X  & X & X  & X & X  & X & X \\
CUSTOM 2       & X & X & X  & C  & X  & C  & AC & X  & X & X  & X & X  & X & X  & X & X  & X & X \\
CUSTOM 3       & X & X & X  & C  & X  & X  & X & C  & AC & X  & X & X  & X & X  & X & X  & X & X \\
CUSTOM 4       & X & X & X  & X  & X  & X  & X & X  & X & X  & X & X  & AC & X  & X & C  & X & C \\
CUSTOM 5       & X & X & X  & X  & X  & X  & X & X  & X & X  & X & X  & X  & X  & X & AC & X & C \\
\hline
\end{tabular}
\end{table}

\end{appendix}
\end{document}